\pdfoutput=1 
\documentclass[conference]{IEEEtran}
\IEEEoverridecommandlockouts
\usepackage{comment}
\usepackage{cite}
\usepackage{amsmath,amssymb,amsfonts}
\usepackage{algorithmic}
\usepackage{graphicx}
\usepackage{textcomp}
\usepackage{xcolor}
\usepackage{footnote}
\usepackage{xspace}
\newcommand{\sys}{\mbox{\textsc{SpecCFI}}\xspace}
\usepackage[hyphens]{url}
\usepackage{pifont} 
\usepackage{multirow}
\usepackage{amssymb}
\usepackage{wasysym}
\usepackage{diagbox}
\usepackage{subcaption}
\usepackage{booktabs}
\usepackage[breaklinks,colorlinks]{hyperref}
\hypersetup{citecolor=blue,linkcolor=blue,urlcolor=black}
\usepackage{xcolor,colortbl}
\usepackage{adjustbox}
\usepackage{array}

\newcolumntype{R}[2]{%
    >{\adjustbox{angle=#1,lap=\width-(#2)}\bgroup}%
    l%
    <{\egroup}%
}
\usepackage{booktabs}
\usepackage{caption}
\newcommand{\cut}[1]{}

\usepackage{relsize}
\newcommand{\cc}[1]{\mbox{\smaller[0.5]\texttt{#1}}}

\newcommand{\PP}[1]{
\vspace{2px}
\noindent{\bf #1}}

\usepackage{color}
\usepackage{bold-extra}
\usepackage{listings}
\definecolor{mblue}{HTML}{01B1F0}
\definecolor{mred}{HTML}{C00000}
\definecolor{mgreen}{HTML}{00B04F}
\definecolor{mpurple}{HTML}{70329F}
\definecolor{dkgrey}{HTML}{77797c}

\lstdefinelanguage{myasm} {
	alsoletter=0123456789,
	keywords=[2]{loop, Train_BTB,Foo,foo,bar,baz,main},
	keywords=[3]{endbr64,cfi_lbl},
	keywords=[4]{L1,L2},
	sensitive=true,
	morecomment=[l]{\#},
	morestring=[b]"
}
\def\BibTeX{{\rm B\kern-.05em{\sc i\kern-.025em b}\kern-.08em
    T\kern-.1667em\lower.7ex\hbox{E}\kern-.125emX}}
\begin{document}

\title{ \sys : Mitigating Spectre Attacks using CFI Informed Speculation\\
%{\footnotesize \textsuperscript}
%\thanks{}
}

%\author{Paper \#258}

\author{\IEEEauthorblockN{Esmaeil Mohammadian Koruyeh\IEEEauthorrefmark{1}, Shirin Haji Amin Shirazi\IEEEauthorrefmark{1},Khaled N. Khasawneh\IEEEauthorrefmark{2},
\\Chengyu Song\IEEEauthorrefmark{1} and Nael Abu-Ghazaleh\IEEEauthorrefmark{1}}
\IEEEauthorblockA{\IEEEauthorrefmark{1}\textit{Computer Science and Engineering Department} \\
\textit{University of California, Riverside}\\
\{emoha004,shaji007,csong,naelag\}@ucr.edu}   \IEEEauthorblockA{\IEEEauthorrefmark{2}\textit{Electrical and Computer Engineering Department} \\
\textit{George Mason University}
    \\\{kkhasawn\}@gmu.edu}
}
%\and
%\and
%\and

\maketitle

\begin{abstract}
Spectre attacks and their many subsequent variants are a new vulnerability class affecting modern CPUs.  The attacks rely on the ability to misguide speculative execution, generally by exploiting the branch prediction structures, to execute a vulnerable code sequence speculatively.  In this paper, we propose to use Control-Flow Integrity (CFI), a security technique used to stop control-flow hijacking attacks, on the committed path, to prevent speculative control-flow from being hijacked to launch the most dangerous variants of the Spectre attacks (Spectre-BTB and Spectre-RSB). Specifically, CFI attempts to constrain the possible targets of an indirect branch to a set of legal targets defined by a pre-calculated control-flow graph (CFG). As CFI is being adopted by commodity software (e.g., Windows and Android) and commodity hardware (e.g., Intel's CET and ARM's BTI), the CFI information becomes readily available through the hardware CFI extensions. With the CFI information, we apply CFI principles to also constrain illegal control-flow during speculative execution.   Specifically, our proposed defense, \sys, ensures that control flow instructions target legal destinations to constrain dangerous speculation on forward control-flow paths (indirect calls and branches).  We augment this protection with a precise speculation-aware hardware stack to constrain speculation on backward control-flow edges (returns). We combine this solution with existing solutions against branch target predictor attacks (Spectre-PHT) to close all known non-vendor-specific Spectre vulnerabilities.  We show that \sys results in small overheads both in terms of performance and additional hardware complexity.
\end{abstract}

%\begin{IEEEkeywords}
%\shirin{???} component, formatting, style, styling, insert
%control-flow integrity, speculative attacks
%\end{IEEEkeywords}

%-------------------------------------------------------------------------------
\section{Introduction}\label{sec:intro}
%-------------------------------------------------------------------------------

The recent Spectre~\cite{spectre} attacks have demonstrated how speculative execution can be exploited to enable disclosure of secret data across isolation boundaries.
Specifically, attackers can misguide the processor to speculatively execute a read instruction with an address under their control.
Although the speculatively read values are not visible to programs through the architectural state, since the misspeculation effects are eventually undone, they can be communicated out using a {\em covert channel}.
Since their introduction, a large number of attacks following the same pattern (temporary read of sensitive data through speculation, followed by disclosure of this data through a covert channel (e.g.,~\cite{contentionchannel,covertchannel})) have been discovered which enable bypassing different permissions using a number of different speculation triggers~\cite{kiriansky2018speculative,ssp2018,koruyeh2018spectre,maisuradze2018ret2spec,weisse2018foreshadow,van2018foreshadow,evtyushkin-18,stecklina2018lazyfp,armv3aspectre,bhattacharyya2019smotherspectre,schwarz2019store}; it is clear that this is a general class of vulnerability that requires deep rethinking of processor architecture.

Since speculation is essential for the performance of modern processors, to mitigate this threat without severely restricting speculation,
some solutions such as InvisiSpec~\cite{yan2018invisispec} and SafeSpec~\cite{safespec} propose separating speculative data from committed data.  Such an approach, rather than attempting to limit speculation, would isolate possible leakage. However, the principle has to be applied to every micro-architectural structure (e.g., cache, TLB, DRAM row buffer), and it is unclear if this approach could prevent leakage through contention, for example, using the functional unit port side-channel~\cite{port-contention,bhattacharyya2019smotherspectre,renderedinsecure}.

Another direction to mitigate this threat is to restrict speculation if a potentially dangerous gadget can be executed speculatively. For example, Intel and AMD suggest inserting serialization instructions like \cc{lfence} to prevent loading potentially secret data~\cite{intel_mitigations,amd}. Because blindly inserting serialization instructions will have the same effect as disabling speculation, thus severely reducing performance~\cite{intelanalysis}, a better solution is to conditionally insert barriers. The MSVC C compiler~\cite{spectre-msvc}, oo7~\cite{oo7}, and Respectre~\cite{Respectre} use static analysis to identify dangerous gadgets and only insert \cc{lfence} before the identified gadgets. Context-Sensitive Fencing~\cite{taram_csf19} dynamically inserts serialization instructions when a load instruction operates on untrusted data (address), but only for Spectre-PHT.

Our observation is that Spectre-like attacks rely on manipulating the processors' prediction structures (see Section~\ref{sec:prediction} for details) to coerce speculation to an attacker-chosen code gadget.
Therefore, these attacks can potentially be defeated more efficiently by identifying and preventing erroneous speculation when the prediction structures produce a wrong prediction.
As a first step towards this direction, we propose \sys, a lightweight solution to prevent the two most dangerous Spectre variants: Spectre-BTB (v2) and Spectre-RSB (v5).
\sys  prevents these attacks by using control-flow integrity (CFI) principles to identify when a prediction is likely erroneous and constrains speculation if it is.

In contrast to traditional CFI, even hardware supported proposals, whose purpose is to prevent illegal control flow within the primary architecturally visible control flow of a program, \sys pushes CFI to the speculation level, where it can be used to determine whether a speculative execution path should be allowed or limited.
Compared to existing solutions against Spectre-BTB and Spectre-RSB, such as the recent microcode update from Intel~\cite{intel_mitigations} and retpoline~\cite{retpoline},
\sys introduces less performance degradation as it still allows correct speculation to proceed,
while these existing solutions blindly ``disable'' all indirect branch prediction.

We also like to argue that defenses against Spectre-BTB and Spectre-RSB serve as the foundation for defense against Spectre-PHT (v1) attacks.
The reason is that serialization instructions can be viewed as a special type of inline reference monitor and, therefore, it is crucial to make sure that these inserted barriers are never bypassed.
However, without protections against Spectre-BTB (forward indirect branches) and Spectre-RSB (returns), attackers can easily bypass the barriers to carry out the attacks~\cite{bhattacharyya2019smotherspectre}.
Furthermore, as demonstrated in return-oriented programming~\cite{shacham07}, by jumping to the middle of an x86 instruction, attackers can use unintended gadgets, in our case speculatively, to launch attacks.
For this reason, we envision \sys being combined with existing solutions against v1 attacks~\cite{taram_csf19,slh,oleksenko2018you} to provide comprehensive protection against Spectre attacks.

The \sys principle can leverage any CFI implementation (e.g., coarse-grained such as Intel's CET~\cite{intel-cet}, or fine-grained such as HAFIX~\cite{davi2015hafix}), with small differences in implementation and leading to the enforcement of the respective version of CFI.   We present our baseline design for forward edge protection in Section~\ref{sec:forward} and backward edge protection in Section~\ref{sec:backedge}.   We investigate two versions of \sys: \sys-base that implements CFI only for speculation, and \sys-full that also supports CFI for the committed control flow (i.e. conventional goal of CFI). 
Section~\ref{sec:eval} evaluates performance and complexity of the design.  We show that \sys-base eliminates dangerous misspeculation (where the predicted target label does not match the destination), without impacting performance.
   
\sys-full incurs an additional small overhead, on par with other hardware CFI implementations~\cite{davi14,davi2015hafix,christoulakis2016hcfi}.  We also analyze the implementation complexity and find that the overhead is modest.

Although some software and hardware solutions have started to appear to defend against this class of attacks, we believe that our solution is elegant along with a number of interesting properties.  We believe that it also combines well with other proposed defenses, such as SafeSpec~\cite{safespec} and InvisSpec~\cite{yan2018invisispec} which limit the speculative side effects once misspeculation occurs, by limiting the opportunities for harmful speculation.   Section~\ref{sec:related} compares \sys to these and other works.

In summary, the contributions of the paper include:
\begin{itemize}
\setlength{\itemsep}{0in}
    \item We present a new defense against Spectre variants that rely on polluting the BTB and RSB, by embedding CFI principles into the branch prediction decisions.
    \item We analyze the security of the proposed designs showing that it protects against all variants of Spectre-BTB (v2) and Spectre-RSB (v5) attacks.  Combined with solutions such as context-sensitive fencing, we believe that we can completely secure the system against Spectre attacks.
    \item We analyze the performance and complexity of \sys, showing that it leads to little overhead.  
    Compared to a defense that prevents speculation around indirect jumps, indirect calls and returns, \sys provides equivalent security yet still avoids the large performance overheads.  The hardware complexity is also negligible.
\end{itemize}
\section{Background}
\label{sec:background}

This section overviews some background: branch predictor structures in modern processors, Spectre attacks, and CFI.

\subsection{Branch prediction and Spectre attacks}
\label{sec:prediction}
Branch prediction is a critical component of modern processors that support speculative out-of-order execution.  When a control flow instruction (branch, call or return) is encountered, the result of the instruction (e.g., whether or not a conditional branch will be taken or what the target value is of an indirect branch or a return) is generally not known at the front end of the pipeline.  As a result, to continue to fill the pipeline and utilize the available resources of the processor, branch prediction is used.  

\begin{figure}[h]
\begin{center}
   \includegraphics[width=0.5\textwidth]{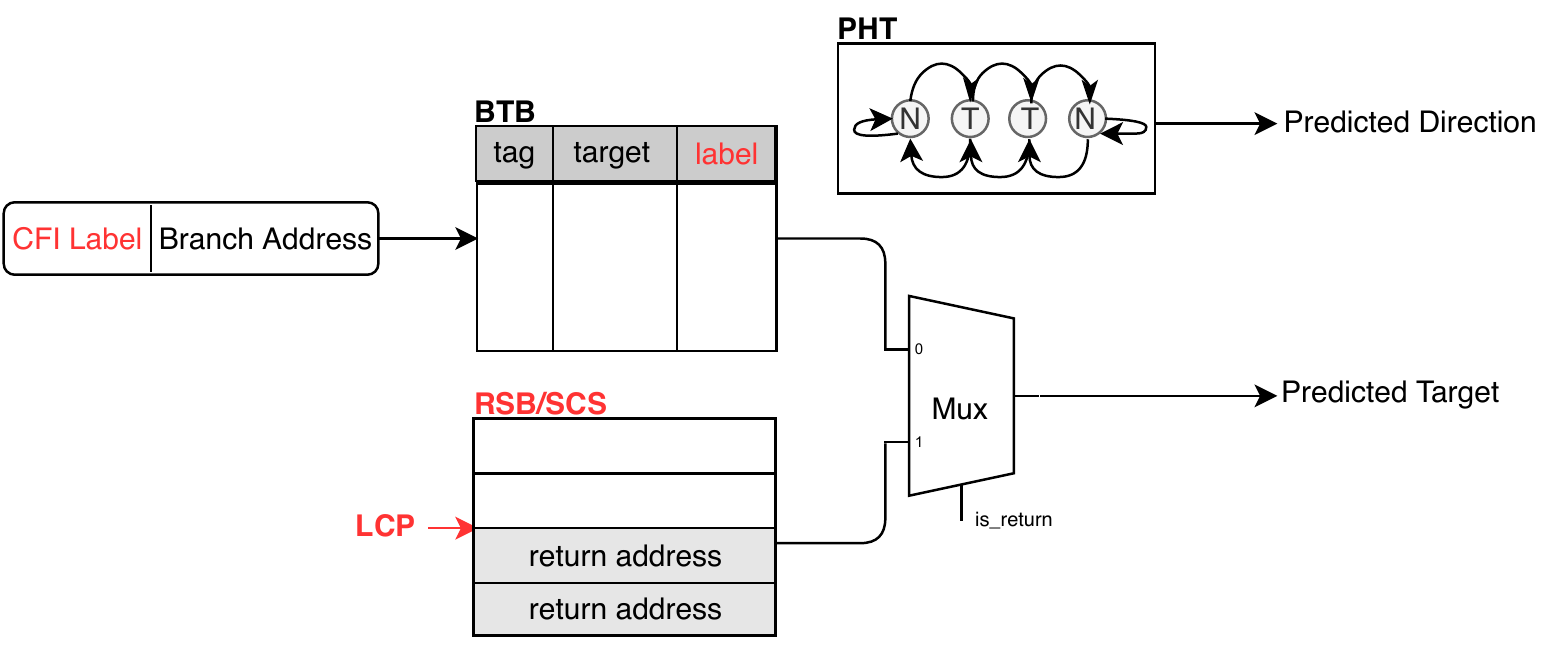}
   \caption{Branch Predictor Unit consists of three different predictors: (1) PHT for conditional branch direction; (2) BTB for indirect branch addresses; and (3) RSB for return addresses.}
   \label{fig:branch_pred}
  \vspace{-1.5em} 
\end{center}
\end{figure}

 Modern processors employ sophisticated predictors (shown in~\autoref{fig:branch_pred}) which typically consist of three components:
\begin{itemize}
\setlength{\itemsep}{0in}
    \item Direction predictor: is responsible for predicting the direction of a conditional branch.  Although a number of implementations have been studied, modern predictors typically implement a two-level context sensitive predictor~\cite{evtyushkin-18}.  The first level is a simple predictor that hashes each branch address to a direction predictor (typically a 2-bit saturating counter). %\shirin{second level?} 
    This predictor is used either when a branch is not being successfully predicted %\shirin{??} 
    or when the predictor has not been trained yet. 
    When the predictor is trained, it typically uses a second prediction algorithm, often a variant of a gshare predictor~\cite{yeh-92}, which uses the global history of a branch in addition to its address to hash to a direction predictor as before.  The advantage is that the same branch can have different predictions based on the control flow path used to reach it.  
    
    \item Target predictor: is used by indirect jump and indirect call instructions which jump to an address held in a register or a memory location, which is unknown at the front end of the pipeline.  This predictor typically uses the hash of the branch address to index a cache holding the branch targets called the branch target buffer (BTB).  BTBs are shared across threads on a virtual core: one value used by a process could be used by another process whose branch has a matching address in the BTB~\cite{evtyushkin-16}.
    %This is the origin of the Spectre v2 injection where one program poisons the BTB with a selected address to cause another program on the same virtual core to use that address when it speculates.
    
    \item The return address stack: Since returns are not well predicted using the BTB, and often follow strict call-return semantics, their target is predicted using a return address stack of fixed size.  When a call instruction executes, the return address is pushed on this hardware stack; if overflow happens, previous entries are overwritten~\cite{koruyeh2018spectre}.  When a return is encountered the top of the stack is popped and used as the return target.  
    %Recently speculation attacks that use the return stack buffer have also been demonstrated~\cite{koruyeh2018spectre,maisuradze2018ret2spec}.
\end{itemize}

\PP{Spectre Attacks}
Spectre attacks exploit the branch and aliasing predictors to fool them to access unauthorized data speculatively~\cite{spectre,kiriansky2018speculative,ssp2018,koruyeh2018spectre,maisuradze2018ret2spec,canella2018systematic,bhattacharyya2019smotherspectre}. The main properties that the attack exploits in speculative execution are:
(1) lazy permission checks on speculation: while instructions are being executed speculatively, the processor will not check the permissions until the commit stage;
(2) Speculative instructions have unintended side-effects on micro-architectural states even if they do not get committed;
and (3) attackers can deliberately mislead execution into attacker-intended gadgets by mistraining branch predictors, and use the previous property to leak sensitive information.  Specifically, an attacker selected gadget is executed speculatively to perform unauthorized access and leak the value through a side-channel~\cite{spectre,armv3aspectre,intelanalysis}.  Based on the prediction structure being attacked, variants of the Spectre attacks that are addressed in this work are shown in~\autoref{tbl:spectre_variants}.  Mitigations for other variants of the Spectre attacks as well as variants of the Meltdown attack have been discussed in detail by Canella et al.~\cite{canella2018systematic}.

\begin{table}[]
\centering
\footnotesize
\caption{Spectre attack variants and their targeted branch prediction components }
\label{tbl:spectre_variants}
\resizebox{0.49\textwidth}{!}{%
\begin{tabular}{l|l}
\hline
\textbf{Spectre}      & \textbf{Element exploited}      \\ \hline
Spectre-PHT (v1)~\cite{spectre}  & Pattern History Table (PHT)  \\
Spectre-PHT (v1.1)~\cite{kiriansky2018speculative}  & Pattern History Table (PHT)  \\
Spectre-BTB (v2)~\cite{spectre}  &  Branch Target Buffer (BTB)  \\
%Spectre v4~\cite{ssp2018}  & Store-to-load Forwarding (STLF)\\
Spectre-RSB (v5)~\cite{koruyeh2018spectre,maisuradze2018ret2spec} & Return Stack Buffer (RSB) \\\hline
\end{tabular}
}%
\end{table}

\subsection{Control-flow Integrity}
Control-flow integrity (CFI)~\cite{cfi,pax-future} is a state-of-the-art solution to mitigate control-flow hijacking attacks. In such attacks, attackers corrupt/overwrite control data (i.e., data that controls indirect control transfer, function pointers and return addresses for instance ) to divert the victim program's execution to carry out attacker-chosen logic, for example, to enable malware or open a backdoor. CFI prevents such attacks by enforcing a basic safety property: \emph{software execution must follow only legal paths within a control-flow graph (CFG) determined ahead of time}~\cite{cfi}. Hence, a CFI mechanism always consists of two components: one that computes the CFG of the program and one that regulates the control transfer while it is executing.

\PP{Constructing CFG.} The security guarantee of a CFI mechanism directly depends on the accuracy of the CFG, which can be constructed through static or dynamic analysis. Coarse-grain CFI mechanisms~\cite{zhang13a,zhang13b} generate the CFG using static analysis: any address-taken function can be a legitimate target for any indirect call; any address taken basic block can be a legit target for any indirect jump; and the address of the next instruction after any call can be a legit target for a return. Although coarse-grained CFI can eliminate most illegal control transfer targets, follow-up research has shown that the CFG used is too permissive/inaccurate that it still allows attacks~\cite{carlini14,goktas2014out}. Fine-grained CFI solutions improve the accuracy of the CFG by incorporating type information~\cite{niu14a,tice14,van2016tough,rap,wang2009fpvalidator}. Unfortunately, the CFG may still allow illegal control transfers~\cite{carlini15,evans15}. More recently, researchers have proposed utilizing run-time information to further improve the precision of the CFG~\cite{niu15,vanderveen15,pittypat}, which can even achieve perfect accuracy~\cite{ucfi} (i.e., one possible target per indirect control transfer site).

\PP{Regulating control-flow.} Once the CFG is calculated, legitimate control transfers can be grouped into equivalence sets. Within the same set, control-flow can be transferred from any source location (e.g., a call site or return site) to any target location (e.g., target function or call site). By assigning each equivalence set a unique ID/label, run-time control-flow can be regulated with a simple check---source label must match destination label. Such checks can be implemented using either software or hardware. Some hardware extensions only support a single label~\cite{kayaalp2012branch,intel-cet,arm-bti} thus can only enforce coarse-grained CFI. Others support multiple labels~\cite{davi14,christoulakis2016hcfi} and fine-grained CFI. Some hardware extensions also include a shadow stack to enforce unique return target~\cite{davi2015hafix,christoulakis2016hcfi,intel-cet,kayaalp2012branch}.

\PP{Adoption.} Because of its effectiveness against control-flow hijacking attacks, CFI has been adopted by both commodity software and hardware. Tice et al.~\cite{tice14} introduced forward-edge CFI to LLVM and GCC in 2014. Android adopted this implementation in 8.1 to protect its media stack and extended the protection in Android 9 to more components and the OS kernel. Microsoft introduced its own CFI implementation, control-flow guard in Visual Studio 2015 and has been utilizing it to protect important OS components, including the web browser. In Windows 10 (V1730), Microsoft extended the protection to the OS kernel and hypervisor (Hyper-V). On the hardware side, Intel introduced Control-flow Enforcement Technology (CET)~\cite{intel-cet} and ARM introduced a similar mechanism,  Branch Target Indicators (BTI), in ARMv8.5-A~\cite{arm-bti}.
%\shirin{would it be a good idea to categorize different CFI implementations and discuss which we are using as one of the reviewer mentioned?}
\cut{ 
\subsection{Motivations}
\label{sec:motivate}

This work is motivated by the observation that Spectre-BTB and Spectre-RSB attacks are similar to traditional control-flow hijacking attacks. The main difference is that the hijacking is achieved through polluting the shared prediction structures (BTB and RSB) instead of corrupting control data stored in memory. Therefore, if the victim software is already protected by a hardware-enforced CFI mechanism, we could easily extend the protection to guard the speculative execution as well.

We note that Meltdown is outside the threat model since it occurs due to speculation within the execution of the same instruction.  Moreover, misspeculation through the direction predictor (which leads to Spectre-PHT) does not result in a control flow violation, since it is simply resulting in the incorrect choice among two legal control flow directions.  Luckily, existing works have already developed protections against Spectre-PHT, primarily by limiting speculation around conditional branches that can lead to dangerous misspeculation~\cite{taram_csf19,slh,Respectre,spectre-msvc}.

}
\section{\sys System Model}
\label{sec:model}

%In this section, we provide an overview of our approach.

This section first overviews the threat model we assume in the paper.   It also  describes the extensions to the Instruction Set Architecture (ISA) to support \sys and the compiler modifications to use them.  %Finally, we describe the compiler modifications to prepare the binary for execution to benefit from the \sys defense.

\subsection{Threat Model}

The main goal of \sys is to prevent attackers from launching branch target injection attacks (i.e., Spectre-BTB and Spectre-RSB).  We assume a strong local adversary model with a shared BTB across different hardware threads (i.e., hyperthread) and protection domains (address space, privilege level, and SGX enclaves).  We assume the RSB is not shared between hardware threads, consistent with existing CPU designs, but it is shared between different protection domains. 
Specifically, we assume adversaries can inject arbitrary branch targets into BTB in an attempt to control the predicted branch target in the victim protection domain.

Meltdown style attacks~\cite{meltdown,van2018foreshadow,Schwarz2019ZombieLoad,ridl,fallout,schwarz2019store} are outside the threat model since they occur due to speculation on the value to be used within the execution of the same instruction; privileged kernel memory~\cite{meltdown}, L1 cache contents~\cite{van2018foreshadow}, fill buffer~\cite{Schwarz2019ZombieLoad}, in-flight data in modern CPUs (for example: Re-Order Buffer and Line Fill Buffers)~\cite{ridl}, and store buffer~\cite{schwarz2019store,fallout}. Moreover, misspeculation through the direction predictor (which leads to Spectre-PHT) does not result in a control flow violation, since both conditional branch directions are legal control flow paths.  Luckily, existing works have already developed protections against Spectre-PHT, primarily by limiting speculation around conditional branches that can lead to dangerous misspeculation~\cite{taram_csf19,oo7,slh,spectre-msvc,Respectre}.  
Similarly, Spectre-STL is out-of-scope but can be mitigated by disabling speculative store bypass~\cite{intel_mitigations,amd-ssbd,arm_css}.
To the best of our knowledge, \sys is the first hardware design that targets the more dangerous Spectre-BTB and Spectre-RSB attacks even when they use different side-channels (e.g., contention-based side-channel in SMT processors ~\cite{bhattacharyya2019smotherspectre}).

We further assume that target software is protected with hardware-enforced CFI, which marks valid indirect control transfer targets (e.g., \cc{ENDBRANCH} in CET).  Although the target software may contain memory vulnerabilities (e.g., buffer overflows) that could be exploited to achieve arbitrary read and write (i.e., the traditional threat model for CFI), such attacks are out-of-scope of this work.

\subsection{Instruction Set Architecture (ISA) Extension}

Most hardware CFI extensions~\cite{davi2015hafix,davi14,christoulakis2016hcfi,intel-cet,arm-bti}
use target labeling to enforce forward-edge CFI, and a shadow stack to enforce backward-edge CFI.
Without the loss of generality, we assume two modifications to the ISA to inform the hardware of the labels from the CFG analysis:

\begin{itemize}
\setlength{\itemsep}{0in}
    \item Extending the indirect \cc{jmp} and \cc{call} instructions to include CFI labels. For coarse-grained CFI enforcement (e.g., Intel CET~\cite{intel-cet} and ARM BTI~\cite{arm-bti}),
    the label at jump and call sites can be omitted.
    %In order to enforce CFI at the hardware level, we extended the ISA by modifying indirect call and jump instructions to embed CFI labels. 
    \item Adding a new instruction to mark legitimate indirect branch targets with corresponding labels.
    For coarse-grained CFI enforcement, the label can be omitted (e.g., the case of Intel CET)
    or collapsed to two labels: one for jump targets and the other for call targets (e.g., the case of ARM BTI).
    %in order to specify the label of valid targets of each indirect branch or call, we added a new instruction containing the label on each valid target based on our CFI analysis called cfi\_lbl. 
    \end{itemize}

\noindent
The shadow stack is generally transparent to the program and will not be directly manipulated.
However, certain language features such as exception handling, \cc{setjmp/longjmp}, require manipulation of the shadow stack.
To support these features, additional instructions are needed, but 
since they do not interact with \sys, we omit their details.
The Intel CET specifications~\cite{intel-cet} provide an example of such instructions.
\autoref{tab:isa} summarizes required ISA changes.

\begin{table}[h!]
\centering
\caption{ISA Extensions to support CFI.}
\begin{tabular}{l l }
\toprule
\textbf{Instruction} & \textbf{Description}\\
\midrule
\texttt{call [dest],label} &  Target class-aware call\\
\texttt{jmp [dest],label} &   Target class-aware jump\\
\texttt{cfi\_lbl} &   Verify CFI integrity\\
\bottomrule
\end{tabular}
\label{tab:isa}
\end{table}
\vspace{-0.8em}%\shirin{added for arxiv}

\subsection{Compiler Modification}

\sys relies on the compiler to mark valid indirect control transfer targets with labels.
Fortunately, because these required modifications are the same as CFI, they are already available as part of commodity compilers.
For example, both LLVM and GCC  include support for (1) software-enforced fine-grained forward-edge CFI~\cite{tice14}, (2) Intel CET, and (3) ARM BTI.
Therefore, \sys requires little or no modifications to the compilers.  \sys is compatible with any label based CFI implementation.

\section{Forward-Edge Defense}
\label{sec:forward}

In this section, we describe the component of \sys responsible for preventing both misspeculation as well as control-flow that breaks CFI on the forward-edge (i.e., on indirect calls and indirect jumps). This defense is responsible for preventing Spectre-BTB (v2) both within the same address space and across different address spaces.  It is also responsible for maintaining CFI integrity on committed instructions (the traditional use of CFI).  

%\reza{Do we need to discuss all target fencing (insert lfences in the target of all indirect branches) as a solution here? or it is enough to mention it in the evalution part?}
\subsection{Preventing Spectre-BTB (within the same address space)}

In this attack, the attacker pollutes the target BTB entry by repeatedly executing an indirect branch in its own address space that hashes into the same entry.  The attacker can use script engines like the JavaScript engine in browsers and the BPF JIT engine in the kernel.  When the victim branch is executed speculatively, the polluted entry will direct the victim to a malicious gadget.  Our goal is to prevent the victim from jumping speculatively to the malicious gadget.

Our first design considers augmenting the BTB to hold a CFI label for the target.  This design extends indirect {\tt call/jmp} instruction execution to update the BTB to add the CFI label of the branch. Later in the speculation path, all indirect calls and jumps are indexed to the BTB to predict their target as before, but with an additional check against the inserted CFI label.  This defense prevents attacker-controlled misspeculation since the label of the attacker's instruction does not match the true target.  For benign programs, such misspeculation is likely to occur only when the BTB is cold (has not been initialized yet), or when branch aliasing causes collisions in the BTB structure.  While these cases should be rare, in both cases the value in the BTB is not the correct target. Limiting such erroneous speculation might result in performance improvement since we do not waste time on fetching instructions from what is likely to be the wrong path.

Since only committed indirect branches update the BTB,  possible targets that may be used by attackers are limited to gadgets starting with a {\tt cfi\_lbl} instruction with an identical label to that of the {\tt call/jmp} instruction’s label.  Note that a label may be shared by multiple locations in the code in CFI, and misspeculation among these locations is still possible (i.e., control flow bending~\cite{carlini14}); as known from CFI solutions, this set is much smaller than the potential targets set without CFI.
%it is highly restricted to the set of targets with a matching CFI label.

\subsection{Preventing Spectre-BTB (cross-address-spaces)}
\begin{figure}[h]
\centering
\begin{tabular}{p{3.5cm} | c}
\hline
\begin{minipage}[t]{3.5cm}
\begin{lstlisting}[language=myasm]
0x09: load rax, 0x25
0x10: call *rax, L1    
...
0x25: cfi_lbl L1
0x26: add rbx,1 
\end{lstlisting}
\end{minipage}
&
\begin{minipage}[t]{3.9cm}
\begin{lstlisting}[language=myasm]
0x09: load rax, 0x50 
0x10: call *rax,L1
...
0x25: load rbx,[secret]
0x50: cfi_lbl L1
\end{lstlisting}
\end{minipage}

\\ \\
\centering \footnotesize{\textbf{\texttt{(Attacker)}}} & \footnotesize{\textbf{\texttt{(Victim)}}}
\\
\hline
\end{tabular}
\caption{Example attack across address spaces}
\label{fig:code1}
\end{figure}
%\begin{figure}[h!]
% \centering  
%  \includegraphics[width=1\linewidth]{figure/code.pdf}
%  \caption{Example attack across address space}\label{fig:cross}
%  \vspace{-1.5em}
%\end{figure}

Storing CFI labels in BTB entries mitigates attacks within the same address space, but not those across address spaces, when attackers pollute the globally shared BTB from another program.  In this case, if attackers know the label used by the victim program (e.g., through offline analysis), they can craft an entry in the BTB with the same label and bypass the protection.  Consider the example in Figure~\ref{fig:code1}.   The attacker inserts L1 and 0x25 in the 0x10 index of BTB, by selecting the label and location of a branch.  
%This entry is valid since the target is annotated with {\tt cfi\_lbl} L1, the same label as the call. 
When the CPU context switches to the victim space, the victim call at location 0x10 is indexed to BTB and uses the BTB entry, inserted by the attacker to predict its target. Since the label matches, the CPU continues speculative execution of the malicious gadget from 0x25, and the attacker successfully redirects the control flow and executes the malicious gadget to reveal the secret.   
\begin{figure}[t!]
 \centering  
  \includegraphics[width=0.9\linewidth]{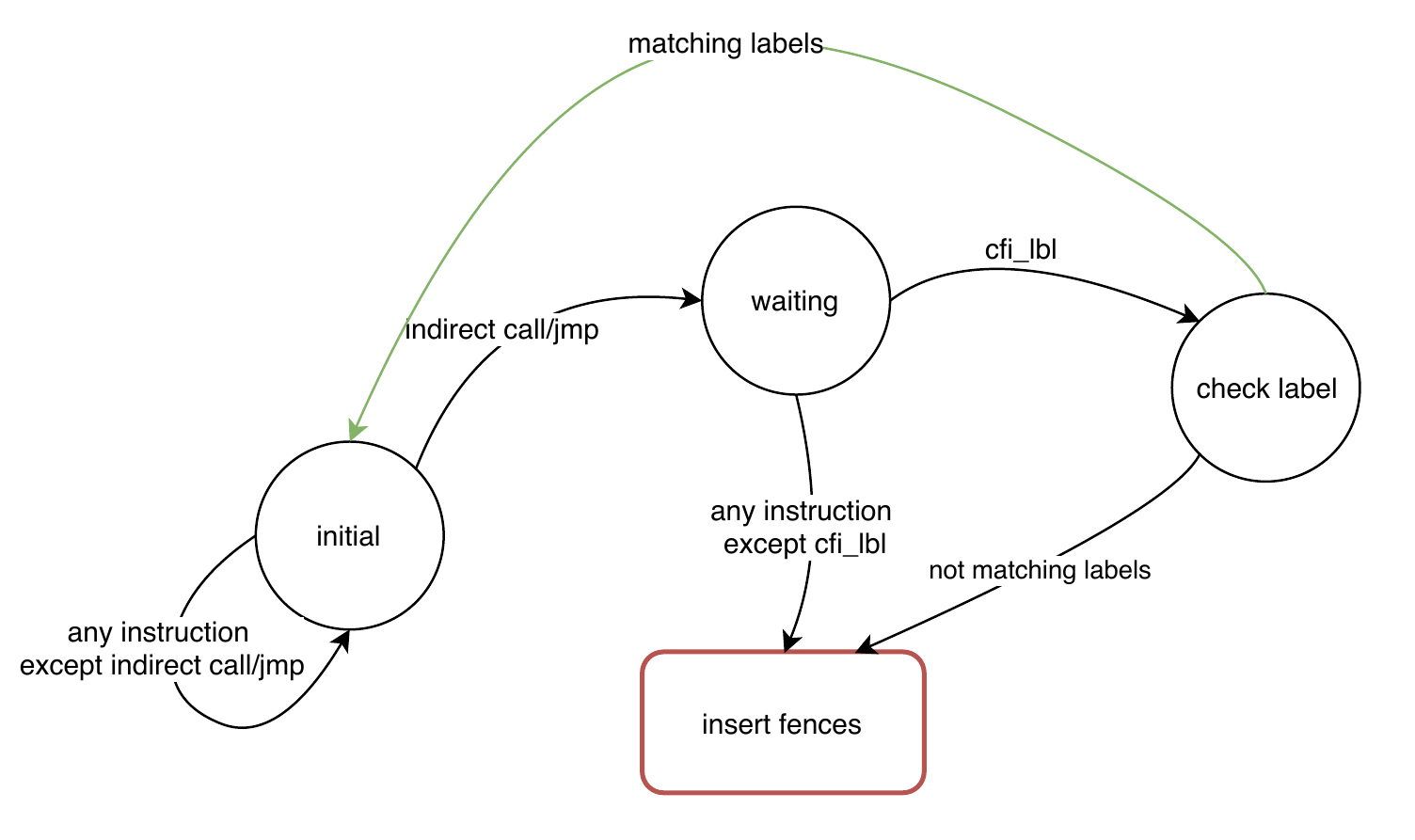}
  \caption{State machine for forward edge protection}\label{fig:state}  
  %\vspace{-2.1em}%\shirin{added for arxiv}
\end{figure}

To prevent cross-address-space attacks, one possibility is to randomize the mapping of addresses to the BTB (e.g., similar to the CASESAR solution for caches~\cite{qureshi-18}) to make it difficult for attackers to guess the label or the location associated with the target branch.  However, as this approach only provides probabilistic guarantees against attacks, we decided to use an alternative implementation that avoids using labels in the BTB.  Specifically, our implementation enforce the CFI check by ensuring that the first speculatively executed instruction after an indirect branch is a legal {\tt cfi\_lbl} instruction with a matching label, guaranteeing that the speculation target is a legal target in the program's Control Flow Graph.  We note that this is the standard implementation of hardware acceleration of CFI.  However, since we are using CFI to constrain speculation (not just the committed instructions), this approach requires pushing the check earlier in the pipeline to the decode stage of the first instruction on the speculative path.  However, as our experimental analysis shows, this change results in negligible impact on performance legal speculation is not delayed.  

  With respect to performance, the two implementations operate differently, but are likely to perform similarly.  The first implementation requires modifications to the critical BTB structure and can potentially slow down the execution pipeline, favoring the target label-checking implementation.  A small disadvantage of the second implementation is that the target instructions have to be speculatively fetched (if not cached) to be able to check the label, which could be avoided if the label-mismatch is detected by the BTB in the first implementation.  %However, this should lead to no loss of performance since these instructions cannot be speculatively executed anyway until they are fetched.
  
The state machine implementing the check in the decode stage of the pipeline is shown in \autoref{fig:state}.  %Unlike the initial design modifying the BTB,  we do not reference the BTB for checking the labels.   
Starting at the initial state, any indirect {\tt call/jmp} instruction in the decode stage sets the {\tt CFI\_REG} register with its own CFI label and causes the CPU to wait for a {\tt cfi\_lbl} instruction. The decode stage makes sure that the next instruction is a \texttt{cfi\_lbl} instruction. This restricts potential gadgets to those starting with a {\tt cfi\_lbl} instruction.  Moreover, the CPU will confirm that the {\tt CFI\_REG} value and the label of the {\tt cfi\_lbl} instruction are equal. In this way, potential gadgets are further restricted to those with a matching label.  When the instruction following the {\tt call/jmp} is not a {\tt cfi\_lbl} instruction or when the label of the {\tt cfi\_lbl} instruction does not match the label of the {\tt call/jmp}, an {\tt lfence} micro-op is inserted into the pipeline to guarantee prevent execution from the wrong speculative path.

\subsection{Enforcing CFI for Committed Instructions}

%The design presented thus far prevents all variants of Spectre-BTB attacks.  
\sys is essentially hardware-supported CFI, but with CFI enforcement during speculation.  Thus, given the similarity in the hardware support to traditional CFI, we also extend the design to support standard CFI to enforce the CFI rules on committed instructions and defend against control flow hijacking attacks.   This support is achieved by enforcing the CFI check during the commit stage of the pipeline: if an indirect {\tt call/jmp} instruction is not followed by a {\tt cfi\_lbl} instruction with a matching label, the CPU raises a CFI violation exception.

\section{Backward-Edge Defense}
\label{sec:backedge}

%\subsection{Overall Design:}
The backward-edge defense component of \sys protects misspeculation on return instructions.  Return instructions typically obtain their predicted addresses from a hardware stack called the Return Stack Buffer (RSB).  The RSB has been shown to be vulnerable to a range of Spectre attacks~\cite{koruyeh2018spectre,maisuradze2018ret2spec}.  To provide protection for the backward-edge, hardware CFI proposals use a Shadow Call Stack (SCS), which is protected from normal memory reads and writes, and can only be manipulated through special instructions~\cite{intel-cet}.  Similar to RSB, the SCS is used to retain the return addresses of previously executed calls.  The differences are: (1) SCS is in memory, so it is saved and restored across context-switch; while RSB is a special cache in the CPU and its content is shared across different context.
%The SCS must be maintained precisely, saved across context switches, and spilling to memory when its capacity is exceeded.  
(2) SCS is only used for CFI enforcement and its size is configurable; while RSB is only used for speculation, and since misspeculation was thought to be only a performance problem, RSB is a best effort structure that is not maintained precisely and has a limited size.

\subsection{Combined Speculation-consistent RSB/SCS: Overview} 

To provide defenses against Spectre-RSB attacks, we combine the traditional RSB and SCS into a unified structure RSB/SCS acting as both RSB and call stack.  Conceptually, RSB in our design can be viewed as the in-processor cache for the in-memory SCS.  We note that this is different from other SCS implementations that retain the RSB separately.  By getting speculation targets from the precisely maintained SCS, consistent with the philosophy of \sys, we move the CFI guarantees to the speculation stage, closing the Spectre-RSB vulnerability.  

The overall design of RSB/SCS has additional requirements from the design of conventional SCS.  Specifically, since we have to be able to use it to obtain speculation targets, it must track additional speculative state without affecting the committed state of the SCS.  We describe the overall design in the remainder of this section.

\begin{figure*}[h]
\begin{center}
   \includegraphics[width=0.65\textwidth]{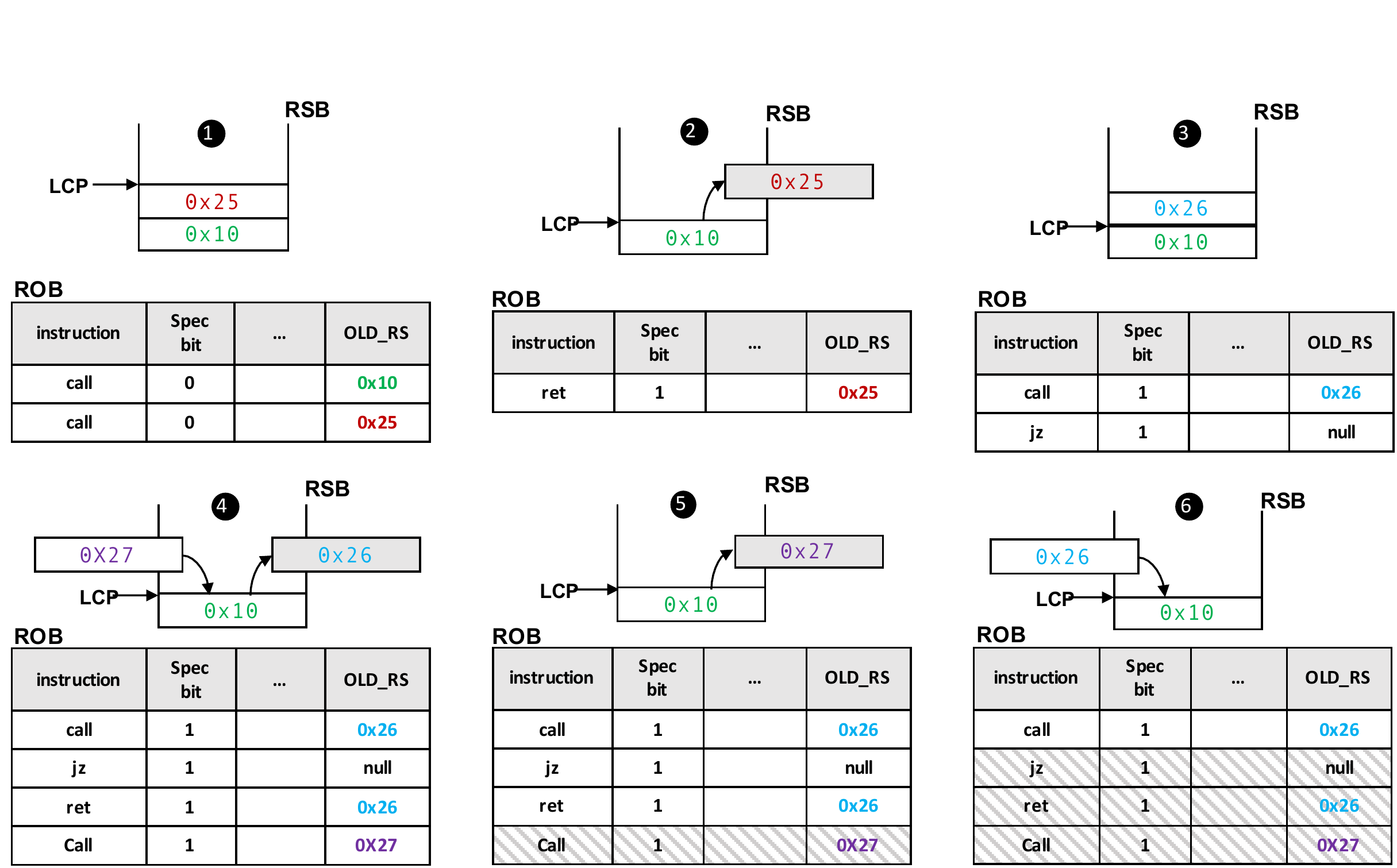}
   \caption{Example of the operation of the combined RSB/SCS}
   \label{fig:ex_rsb_scs}
% \vspace{-1.5em} 
\end{center}
\end{figure*}

When a context switch occurs, the committed RSB/SCS entries must be saved such that they can be restored when the program runs again.  
%\par
To be able to keep the state of this structure consistent, we extend the reorder buffer (ROB, which is the structure in the CPU used to track speculative instructions and their register values before they commit) to track this state.
Specifically, we add a logical register {\tt OLD\_RS} which (is subject to renaming and) holds the return address that is pushed to the RSB/SCS by a call instruction, or popped by a return instruction from the RSB/SCS.
%\par
In addition, we keep track of a pointer to the last committed entry (LCP) of the RSB/SCS so as to save and restore the state of committed entries in this structure in the case of context switch or a spill overflow to memory.
%\par
At the decode stage, If the instruction is a call, the next address is ``speculatively'' pushed to the RSB/SCS structure. When this instruction commits, the LCP is updated to point to the last committed entry.
%\par
If the instruction is decoded as a return it ``speculatively'' pops a return value from the RSB/SCS structure into {\tt OLD\_RS} (without changing LCP) and sets the program counter to this address.  To support conventional CFI, when the return instruction reaches the commit stage, the value of the {\tt OLD\_RS} register is compared with the top of the traditional software stack. If these two values do not match, a CFI violation exception is raised. %Otherwise, this return instruction gets committed, and the LCP is decreased by 1 to point to the next committed entry in the RSB/SCS. 

We considered the need to provision the stack with additional ports since it is used not only to serve committed instructions, but also to handle speculative calls and returns. However, we found that additional ports do not result in performance benefits because the speculative SCS state is held primarily in the port-rich reorder buffer.  
When the in-processor cache (RSB) overflows or the current thread is about to be swapped out, we spill it over to the hardware-protected in-memory SCS.
%CSB resides in a protected memory region, inaccessible to the user program (in the case of enclaves, CSB can be made accessible to the OS).   CSB is a structure where the committed entries are saved if they exceed the capacity of the hardware stack, as well as when a context switch occurs.  
When the RSB underflows or a new thread is swapped in, we load entries from the SCS.
We did not explore optimization to prefetch values from the SCS when RSB is close to empty, or to push some values proactively to memory when RSB gets close to full.

%\begin{figure}[h!]
%\centering
%\footnotesize
%\vspace{-1em}

%\begin{minted}{gas}

%0x09    call func1;
%0x10    ...
%0x24    func1:
%        call func2;
%0x25    call func3;
%0x26    call func4;
%0x27    ...
%...
%0x36    func2:
%        ret;
%...
%%0x74    func3:
%%        jz 0x86;
%...
%0x86    ret;
%\end{minted}

%\vspace{-1em}
%\caption{SpectreRSB basic attack example}
%\label{fig:attack1b}
%\vspace{-1em}
%\end{figure}  

\subsection{Misprediction Recovery}

Every {\tt ret} instruction utilizes the RSB/SCS to predict its jump target. Since the state of RSB/SCS is modified by speculative call and ret instructions, in case of misspeculation, the CPU has to recover the correct state of the structure.

When misspeculation is detected, we need to flush all the speculated instructions from the pipeline. As a part of this process, we have to annul all the corresponding entries from the ROB. During annulment, for every call or return instruction, we not only remove the ROB entry but also update the RSB/SCS to preserve the consistent state of the structure. If the instruction is a call, the top of the RSB/SCS is be popped. In the case of a {\tt ret} instruction, the value of {\tt OLD\_RS} will be pushed back to the RSB/SCS.    

\subsection{RSB/SCS Work Flow}
To clarify how this structure works, we step through the example code sample presented in Figure~\autoref{fig:rsb_ex}. 

\begin{figure}[h!]
\setlength{\textfloatsep}{-0.8\baselineskip plus 0.2\baselineskip minus 0.2\baselineskip}
\centering
\begin{lstlisting}[language=myasm]
     0x09:       call Function1;
     <@\textcolor{mgreen}{\textbf{0x10}}@>: 
     0x24:       Function1:
                        call Function2;
     <@\textcolor{mred}{\textbf{0x25}}@>:              call Function3;
     <@\textcolor{mblue}{\textbf{0x26}}@>:              call Function4;
     <@\textcolor{mpurple}{\textbf{0x27}}@>: 
     0x36:       Function2:
                        ret;
     0x74:       Function3:
            	        jz 0x86; 
     0x86:       ret;
     
     
\end{lstlisting}
\caption{Code sample to illustrate the operation of RSB/SCS}
\label{fig:rsb_ex}
\end{figure}
Assume both calls to \emph{function1} and  \emph{function2} have pushed their return values to the RSB/SCS. By committing these instructions at \ding{202}, the LCP is updated to point to the last committed value and then the corresponding entries are evicted from ROB. In the second step \ding{203}, the return instruction from the first call is being executed speculatively, saving the return address in the ROB, and eventually getting committed. The following speculative call to  \emph{function3} at \ding{204}, will push its return address to RSB/SCS. At step \ding{205}, the execution of the return instruction and the following call to  \emph{function4}, change the RSB/SCS state. Assume that a misspeculation on the {\tt jz} instruction has been detected at \ding{206} and every instruction executed after the branch has to be flushed. Therefore, the recovery process starts annulling instructions from the last entry in ROB until the misspeculated instruction has been reached. Annulling the last call in the ROB at \ding{206}, the value at the top of RSB/SCS is popped and at \ding{207}, annulling the return, the OLD\_RES value of the instruction saved in ROB is pushed back to the RSB/SCS to reset the state to the previous state before the misspeculation.

\subsection{Preventing RSB Poisoning}

Since the RSB/SCS is not shared between different threads and preserved across context switches, the attacker is not able to poison this structure.  Although we allow special instructions to manipulate the SCS to take care of cases such as {\tt setjmp/longjmp}, we assume these instructions are only available to code within the trusted computing base to prevent them from being abused to arbitrarily manipulate the RSB/SCS (which is not a Spectre vulnerability).

\section{Security Analysis}
\label{sec:security}

In this section, we analyze whether \sys can achieve its primary security goal:
{preventing attackers from exploiting branch target injection to ultimately launch Spectre attacks.}

\subsection{Guarantees against Branch Target Injection}
Branch target injection attacks target two prediction components: the branch target buffer (BTB) and the return stack buffer (RSB). Similar to CFI, %where the defense does not prevent attackers from modifying control data (e.g., function pointers and return addresses) but aims to prevent attackers from arbitrarily altering the control-flow to execution code gadget they want;
\sys does not prevent such injections: we assume attackers can still insert arbitrary targets into the BTB, for example by executing branches inside their own protection domain~\cite{evtyushkin-16}.
What \sys guarantees is that if the injected target is not a valid indirect control transfer target in the victim protection domain, then the injected prediction target will not be executed speculatively,
i.e., they cannot speculatively execute arbitrary code gadgets.
For RSB, \sys essentially converts it into a precise shadow call stack (SCS) and maintains it across context switches,
such that both in-address-space injection and cross-address-space injection are no longer possible.

\noindent 
\emph{Impact of Imprecise CFG:}
One weakness of static CFG construction is imprecision, leading to having multiple possible targets with the same label.  This ambiguity may still allow attackers to launch attacks using permitted function-level gadgets~\cite{burow2016,evans15,carlini15,schuster2015counterfeit}.
Since \sys also relies on the CFI analysis to provide valid targets for forward-edge indirect control transfer, it also inherits the same limitation: \emph{mis-prediction is still possible to any of the targets sharing a valid label}.  Since \sys is compatible with any label based CFI, it can benefit from improvements in CFI systems that are increasing the precision in tracking the legal control flow.
%However, some of the unique characteristics of Spectre attacks  this weakness does not pose significant security threat.
%We will discuss the details in the next subsection.

\subsection{Incorporating Defense against Spectre-PHT}
\sys on its own can only mitigate Spectre-BTB and Spectre-RSB attacks.
In this subsection, we discuss how \sys can be (and \emph{should} be) combined with Spectre-PHT defenses to complete the defense against known Spectre variants.
In particular, to defend against Spectre-PHT attacks, researchers have proposed code analysis techniques~\cite{spectre-msvc,oo7,Respectre} to (1) identify dangerous code gadgets that can be used to leak information and (2) conditionally insert serialization instructions (e.g., \texttt{lfence}) to prevent these dangerous code gadgets from being executed speculatively.
One tricky part of such analysis is that, although on the committed path, direct control transfer is always correct; during speculation, even direct control transfer can be wrong. As a simple example, consider a direct call behind a conditional branch: if the prediction on the conditional branch is wrong, then the following direct call is also wrong.
For this reason, when analyzing the code to identify potential dangerous gadgets for Spectre-like attacks, one must perform inter-procedural analysis (for both direct and indirect calls) to account for gadgets that may span across function calls.
The unique opportunity here is that, if the static analysis to identify and eliminate Spectre gadgets uses the same CFG for CFI enforcement, then malicious gadgets at the beginning of function should already be eliminated.
As a result, when combined with such defenses, even if \sys allows misspeculation due to imprecise CFG, the wrong target cannot be used to launch attacks, because the gadgets have already been eliminated.

At the same time, defenses against Spectre-PHT attacks have to use {\sys}-like techniques to be sound.
The reason is the same reason inline reference monitors like Software Fault Isolation~\cite{mccamant2006evaluating,yee2009native} have to enforce some control-flow regulation---if attackers can hijack the control-flow to arbitrary locations, then they can easily bypass the inserted checks and bypass the protection.  This is especially dangerous to variable length ISA like x86 where attackers can jump to the middle of an instruction to find unintended instructions forming exploitable gadgets.
Similarly, \sys provides the same runtime guarantee to Spectre-PHT defenses: by enforcing that even speculative control-flow cannot deviate from the CFG used in static analysis, \emph{the code being analyzed and instrumented will be the same as that executed}.

\subsection{Comparison to Intel CET}

A few days before the submission of this paper, Intel published a new specification of its CET~\cite{intel-cet} extensions.  The new specification includes a paragraph (section 3.8) indicating their plans to include a check that an indirect branch executed speculatively targets a legal \verb+Branch_end+ target.  Intel suggested this solution, which is essentially the configuration of \sys using CET as the CFI implementation, concurrently with our work.  

We believe that Intel's interest in this solution validates it practicality as a defense against transient speculation attacks.   While the updated CET specifications document describes only the general idea, our work contributes a reference implementation and assessment of both the performance and security of the solution.  In addition, \sys provides substantial security advantages over the new CET, including:
\begin{itemize}
    \item Backward edge protection using the speculation aware shadow stack.  While Intel CET uses a shadow stack to protect the backward edge for committed instructions, the specifications describe no plans to use it for limiting speculation.  It is not trivial to extend the shadow stack to track the speculative state, as we describe in Section~\ref{sec:backedge}.  

    \item Generalized CFI protection and limiting control flow bending.  CET only enforces that control flow (whether committed or, in the new specifications, speculative) happens to the start of a legal basic block.  As a result, it allows arbitrary control flow bending~\cite{carlini15}, which does not meaningfully restrict the attack opportunities.  In contrast, \sys admits any CFI implementation, which can substantially shrink the control bending attack possibilities.  Specifically, from a given indirect control flow instruction, only the gadgets with matching CFI label are reachable.  State-of-the-art CFI systems such as PathArmor/Context Sensitive CFI can be supported~\cite{vanderveen15} substantially limiting the control flow opportunities.  In particular, we intend to explore supporting uCFI~\cite{ucfi} in our future work, leaving no control flow bending opportunities available.
\end{itemize}

\section{Performance and Complexity Evaluation}
\label{sec:eval}

In this section, we evaluate \sys in terms of  performance and hardware complexity. All performance experiments were conducted using the MARSSx86 (Micro Architectural and System Simulator for x86)~\cite{marssx86}, a widely used cycle accurate simulator. MARSSx86 is built using PTLsim~\cite{ptlsim} and does a full system simulation (including the OS) on top of the QEMU~\cite{qemu} emulator. First, we configured MARSSx86 to simulate an Intel Skylake processor; configurations are shown in~\autoref{tbl:cpu_config}. We then integrated \sys into the simulator to model all new operations realistically and in full details, in order to retain hardware faithful cycle accurate modeling of the extended processor pipeline. 

\begin{table}[t]
\centering
\caption{Configuration of the simulated CPU}
\label{tbl:cpu_config}
\resizebox{0.49\textwidth}{!}{
\begin{tabular}{ll}
\hline
\textbf{Parameter} & \textbf{Configuration}               \\\hline
CPU        & SkyLake \\
Issue       & 6-way issue\\
IQ          & 96-entry Issue Queue \\
Commit         & Up to 6 Micro-Ops/cycle    \\
ROB        & 224-entry Reorder Buffer    \\
iTLB        & 64-entry instructions Translation Lookaside Buffer\\
dTLB        & 64-entry data Translation Lookaside Buffer \\
LDQ        & 72-entry  Load Queue\\
STQ        & 56-entry  Store Queue\\ 
RSB         & 16-entry Return Stack Buffer \\
I-Cache         & 32 KB, 8-way, 64B line, 4 cycle hit \\
D-Cache          & 32 KB, 8-way, 64B line, 4 cycle hit \\\hline
\end{tabular}%
}\vspace{-1.5em}%\shirin{for arxiv}
\end{table}

\subsection{Performance Evaluation}

We use the SPEC2017 benchmarks~\cite{spec2017} for evaluation, which is a standard benchmark suite used to evaluate the impact of processor modification on a range of representative applications that exhibit a range of different behaviors. All benchmarks were compiled using an LLVM compiler that is modified to mark valid indirect control transfer targets with labels. %\reza{: do we need to mention intel cet instead of modified llvm?!}
 Unfortunately, since there is no official LLVM front-end for FORTRAN~\cite{llvm_test}, we were not able to compile 8 out of the 23 SPEC2017 benchmarks as they contain FORTRAN code. 

One option to prevent Spectre attacks is to insert fences to stop speculation around indirect control flow instructions. In order to evaluate \sys performance, we compare it against the following design points:
\begin{itemize}
\setlength{\itemsep}{0in}
    \item Baseline:  this is the case of an unmodified unprotected machine.  Specifically, we compile and run the SPEC2017 benchmarks using unmodified version of LLVM compiler and MARSSx86 simulator. In all of our experiments, we use the Instructions committed Per Cycle (IPC), a common metric for evaluating the performance of processors, to report performance.  The IPC values of the defenses are normalized to this baseline implementation without defenses; thus, a higher normalized value than 1 indicates better than baseline performance. 
    \item \emph{Retpoline-style software fencing}: we implement a system adding fences to indirect branches using software. The compiler is modified to substitute all the indirect branches and return instructions with a sequence of instructions which ensure that the target of the branches are resolved before any following instruction that might touch the cache (i.e, load) are issued. For protecting the forward edges (i.e. indirect call and jumps) This is done by converting each indirect call to the three following instructions: \ding{202} a \texttt{load} preparing the value of the target register/memory, \ding{203} an \texttt{lfence} making sure that no future load is issued before the branch is resolved and \ding{204} the actual call to the address specified in the target register. Taking the same approach for securing backward edges (i.e. returns) we substitute any \texttt{ret} instruction with a sequence of \ding{202} a \texttt{pop} from top of the software stack to the target register, \ding{203} an \texttt{lfence} making sure to stop the speculation before the actual target of \texttt{ret} resolved and \ding{204} a \texttt{jmp} to transfer the control to the target.
    Conceptually, this solution is similar to the Retpoline defense~\cite{retpoline} which essentially replaces speculation on indirect branches with an empty stall gadget. Different from Retpoline, we also insert the fences for returns (Retpoline does not protect returns, and leaves the code vulnerable to Spectre-RSB attacks).  
    
    This software approach has the advantage of not modifying the underlying hardware but imposes a noticeable overhead in the number of instructions and code size. 
    \item \emph{All Target Fencing}: In this approach, we show one implementation with an \texttt{lfence}, inserted in hardware, at target of each indirect branch and return (the all target fencing) since such a defense is possible without CFI.  This is done by detecting every indirect call, jump, or return in the decode stage of the pipeline and inserting an \texttt{lfence} at target of them to make sure that the branch is resolved before issuing further instructions. 
\end{itemize}

\begin{figure}[t]
   \includegraphics[width=\columnwidth]{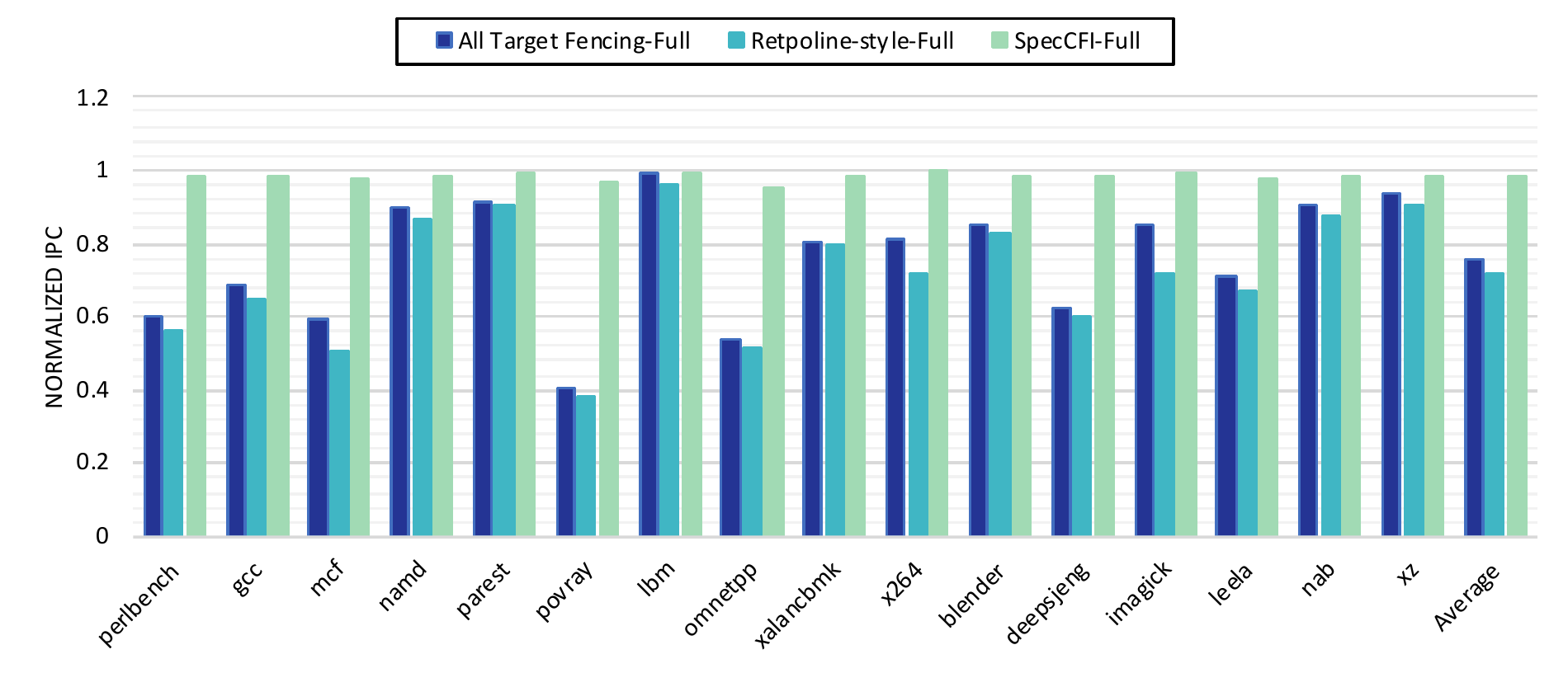}
   \caption{Performance Impact }
   \label{fig:performance}
  \vspace{-1.5em} 
\end{figure}
 %We evaluate the performance of \sys  by comparing performance parameters such as IPC ovearhead between these defenses. We show that \sys has significantly less overhead than all the other approaches. 
The implementations discussed above  prevent speculation by inserting \texttt{lfence} into the pipeline.  \sys offers a more intelligent and targeted way of using fences for securing forward edges (as discussed in Section~\ref{sec:forward}), as well as a new method for making backward edges secure (as explained in Section~\ref{sec:backedge}).
To study the effect of different serializing instruction we use two different types of \texttt{lfence} instructions in our experiments:
\begin{itemize}
\setlength{\itemsep}{0in}
    \item \emph{Strict} \texttt{lfences}, are highly restrictive and prevent any instruction to pass through them until the fence retires~\cite{taram_csf19}. This type of fences impose high overhead to the system. All the x86 serialization instructions including the \texttt{lfence} we use in our experiment, categorize as strict fences.
    \item \emph{Relaxed} \texttt{lfences}, only stop certain types of instructions until the fence gets retired~\cite{taram_csf19}, while letting the others through. For example, LSQ-LFENCE~\cite{taram_csf19}, prevents any subsequent load instruction from being issued speculatively out of the load/store queue but allows any other instruction to pass it. LSQ-LFENCEs are secure against Spectre because they prevent the speculative loads, and have the advantage of letting speculation on other types of instructions proceed, substantially reducing the performance impact.% in our experiment which harms the performance much less compared to strict lfence.
\end{itemize}
%\reza{shoud be updated based on new evaluation results}

\autoref{fig:performance} shows the performance overhead of {\sys-full} (securing both forward and backward edges) in comparison to the \emph{All Target Fencing} and \emph{Retpoline-style software fencing} approaches. We note that in general, inserting serializing instructions (e.g, \texttt{lfence}) in the target of every indirect branch is expensive, imposing performance overhead of 39\% and 48\% on average for \emph{All Target Fencing} and \emph{Retpoline style}  respectively. Using \sys, by inserting \texttt{lfence} only when the CFI check fails, the number of inserted \texttt{lfence} drops significantly thus reducing the performance overhead to less than \emph{1.9\%} on average.

\begin{figure}[t]
\begin{center}
   \includegraphics[width=0.98\columnwidth]{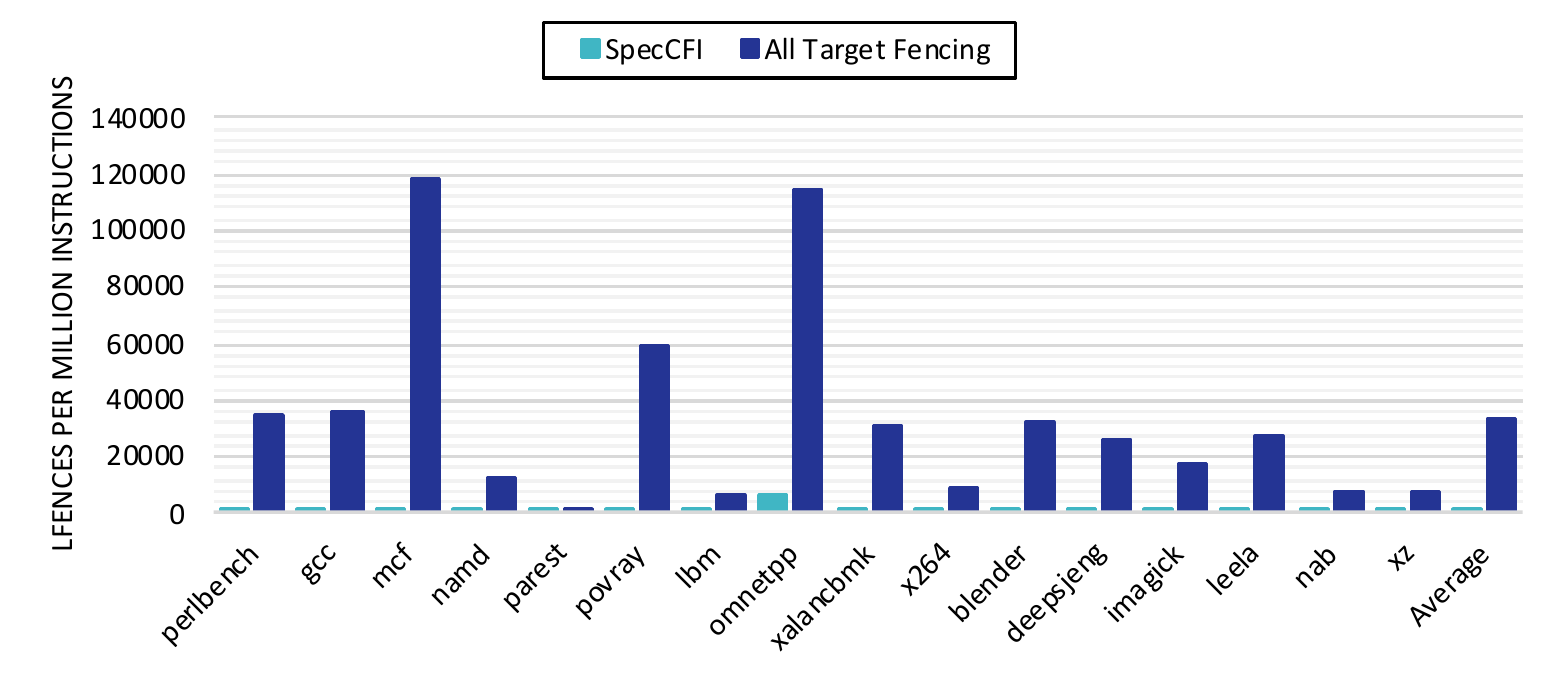}
   \caption{Number of lfences inserted by different defenses}
   \label{fig:lfence}
  \vspace{-1.5em} 
\end{center}
\end{figure}

To illustrate the reason behind the performance reduction in the different approaches, we study the number of \texttt{lfence} instructions inserted in each approach in~\autoref{fig:lfence}.  Note that benchmarks such as \texttt{mcf} and \texttt{omnet}, are C++ benchmarks which use a large number of indirect branches due to the common use of virtual function calls and function pointers.  As a result, this leads to a large number of \texttt{lfence} being inserted into the pipeline, and to a substantial performance impact compared to the baseline implementation.  The only exception to this trend is \texttt{Provay} which suffers the highest overhead for all the defenses but does not have huge number of \texttt{lfence} compared to the other benchmarks.  Looking more closely at this benchmark, we found out that it is a memory intensive benchmark with the highest number of load and store micro-ops among all the benchmarks.  Intel manuals~\cite{intel-manual} indicate that an \texttt{lfence} is committed only  when there is no preceding outstanding store. Thus, for this benchmark, each \texttt{lfence} instruction remains active for a longer period of time until it gets committed which explains the high performance impact.
It is also worth mentioning that unlike the \textit{All target fencing} and \textit{Retpoline-style} which insert \texttt{lfence} for each indirect branch, the   \texttt{lfence} instructions for \sys occur due to mis-prediction detected as a label mismatch causing the insertion of the \texttt{lfence}. This means that the higher the rate of mis-prediction, the more \texttt{lfence} instructions are inserted.

\begin{figure}[t]
\begin{center}
   \includegraphics[width=\columnwidth]{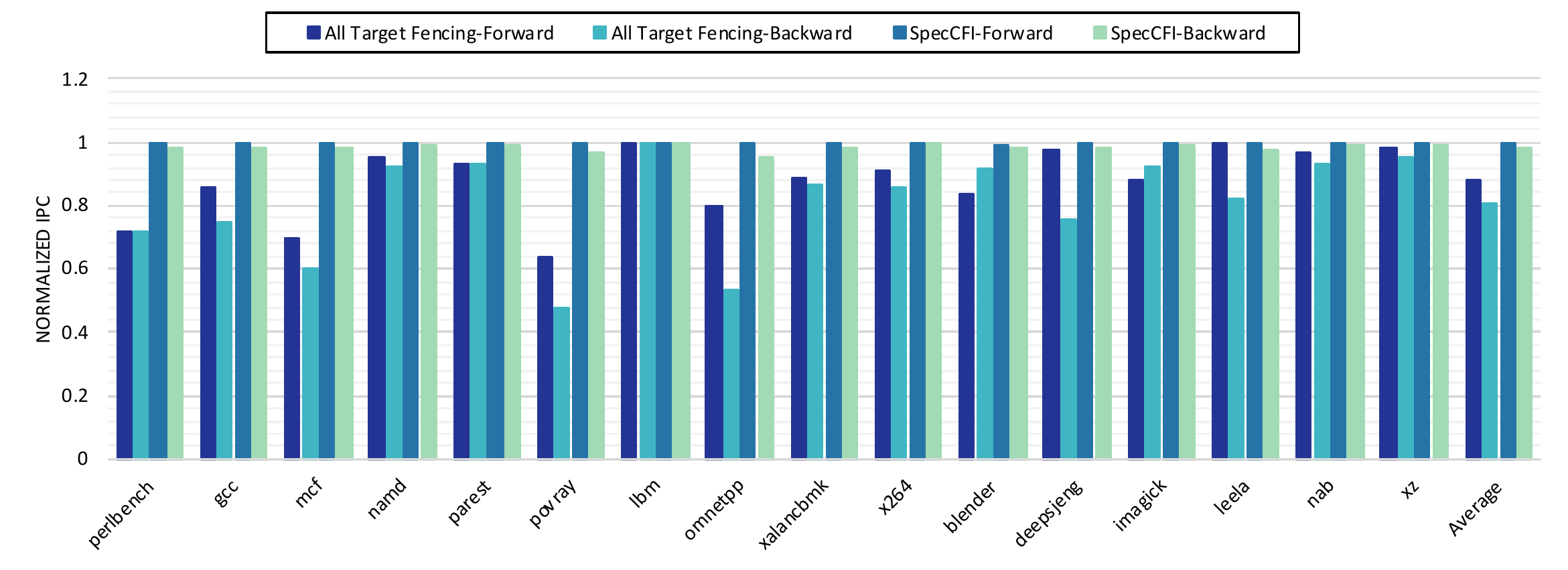}
   \caption{Overhead breakdown for forward and backward edge}
   \label{fig:breakdown}
  \vspace{-2em} 
\end{center}
\end{figure}

In~\autoref{fig:breakdown}, we study the effect of securing the forward and backward edges separately since they use separate mechanisms for protection. Note that in \textit{Retpoline-style}, all return instructions are converted to a sequence of instructions terminating with a \texttt{jmp}, meaning that there is no remaining \texttt{ret} instruction (i.e. backward-edge) in the code compiled in this setting. Therefore, the overhead measured as the overhead of \textit{Retpoline-style}-full is equivalent to only \textit{Retpoline-style}-forward overhead and the overhead on the backward-edge is zero. The results from the breakdown show that as expected, the overhead in general increases with the number of indirect branches in \textit{All Target Fencing}. As for \sys, the overhead caused from forward edge defense is typically low: the overhead is incurred only on CFI mismatches which indicate misprediction of the branches. Therefore, the major part of the \sys overhead is the overhead of \sys-full on the backward-edge which is associated with maintaining the RSB/SCS hardware structure. It is important to consider that this maintenance effort also includes procedures to make sure the committed path is secure and therefore only a portion of this overhead is associated with defense against Spectre attacks. 
%In contrast, \sys enables us to stop speculation only on a misprediction, performing similar to the baseline processor.

\begin{figure}[t]
\begin{center}
   \includegraphics[width=\columnwidth]{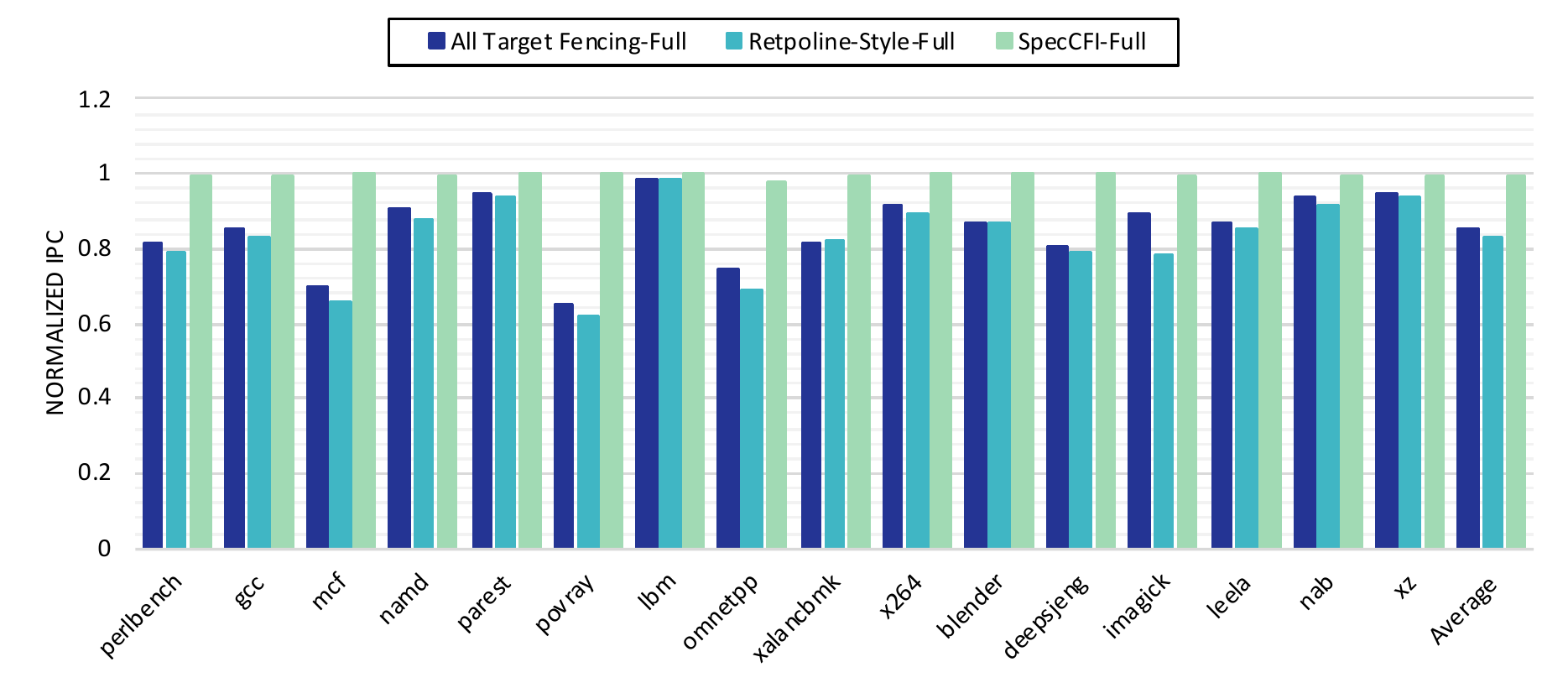}
   \caption{Performance using relaxed fences}
   \label{fig:relaxed}
  \vspace{-1.5em} 
\end{center}
\end{figure}

Since strict \texttt{lfence} imposes a higher overhead on the system and relaxed \texttt{lfence} provides the same security guarantee with lower overhead, we implemented all discussed defenses with relaxed \texttt{lfence} as well to study the differences in overhead.  \autoref{fig:relaxed} examines the effect of relaxed \texttt{lfence}.  The results show that the overhead caused by strict \texttt{lfence} is much higher than that of relaxed \texttt{lfence}.  Also as expected, using strict instead of relaxed causes far more performance degradation when the benchmark is memory intensive (i.e., has a lot of stores in this case).  Our results show that just by changing the type of the \texttt{lfence} from strict to relaxed, the average overhead drops down from 48.9\% to 22.6\% for \textit{Retpoline-style} and from 39.9\% to 18.82\% for \textit{All Target Fencing}.  However, these overheads are still substantially higher than those of \sys.

\subsection{Hardware Implementation Overhead}

To estimate the hardware overheads of \sys, we implemented the primary hardware structures and integrated them within an open core to estimate the area and timing overhead. Specifically, the implementation consists of adding two \texttt{CFI\_REG} registers in two locations of the pipeline: (1) decode stage, to support detecting CFI violations for speculative instructions and (2) commit stage, to support detecting CFI violations for committed instructions. Since \texttt{CFI\_REG} is used to store the CFI labels its size should be the same as the maximum CFI label size (32-bits for our design). Furthermore, we need to add two comparators; one in decode and one in commit stage of the pipeline. These comparators will be used by \texttt{cfi\_lbl} instruction to compare its label to the \texttt{CFI\_REG} (todetect violations).

Additionally, \sys needs a \texttt{LCP} register to point to the last entry of the RSB/SCS from a committed call, used to distinguish between entries from speculative and committed instructions.  Since RSB/SCS has 16 in-processor cache entries, the \texttt{LCP} size is 4-bit. Moreover, at two stages of the pipeline, new entries can be added to the RSB/SCS: (1) while executing call instruction and (2) load the preserved RSB/SCS entries from memory in case of underflow. Therefore, we had to update the number of write ports from 1 to 2. The same thing applies to the number of read ports, as we may use RSB/SCS to fetch next instruction while spilling over to memory in case of RSB/SCS overflow. In addition, to preserve the correct behaviour of RSB/SCS, we provided two \texttt{LCP} update mechanisms: (1) -/+1: for regular push/pop operations and (2) -/+4: for handling overflow and underflow of the structure.  The cost of the RSB/SCS itself did not lead to a noticeable increase in complexity or area.

To measure the impact of \sys implementation on power, area, and cycle time, we modified the open source processor (AO486)~\cite{ao486} to include \sys design using Verilog. To synthesize the implementation of integrating \sys to the processor on a DE2-115 FPGA board~\cite{altera-DE2} we used Quartus 2 17.1 software. The results shown in~\autoref{tbl:hardware_overhead} prove that \sys indeed has low implementation complexity. In terms of power, there is a 0.4\% increase in core dynamic and static power. Although it is difficult to measure power accurately, we applied the power analysis tool provided by Quartus to measure power after synthesis to get more accurate results. In terms of area, there is a 0.1\% increase in total logic elements. Moreover, since \sys design is simple, it fits within the optimized frequency of the core. Thus, it has no effect on cycle time. The AO486 processor is an implementation of the 80486 ISA using a 32-bit in-order pipeline. Thus, these results are relative to the small pipelined core; the overheads will be much smaller if compared to a modern out-of-order superscalar core.

\begin{table}[t]
\centering
\small
\caption{\sys hardware implementation overhead after adding it to the AO486 open-core}
\label{tbl:hardware_overhead}
\resizebox{0.48\textwidth}{!}{%
\begin{tabular}{l|cccc}
 %\hline
                                           &
             \begin{tabular}[c]{@{}c@{}}\textbf{Static power}\end{tabular}  &
             \begin{tabular}[c]{@{}c@{}}\textbf{Dynamic power}\end{tabular}  &
             \begin{tabular}[c]{@{}c@{}}\textbf{Area}\end{tabular}  &
             \begin{tabular}[c]{@{}c@{}}\textbf{Cycle time}\end{tabular}   \\ \hline
 \sys         &  0.4\% &  0.4\%    & 0.1\%     &  0.0\%    \\ \hline
\end{tabular}
%\smallskip
}\vspace{-1.5em}
\end{table} 

\subsection{Empirical Security Evaluation}

\subsubsection{Against real exploits}
To verify our analysis, we evaluated the effectiveness of \sys against real-world exploits.
We ran previously disclosed Spectre-BTB~\cite{spectre}, Spectre-RSB~\cite{koruyeh2018spectre}, and SMoTHerSpecter~\cite{bhattacharyya2019smotherspectre} PoC inside the emulator.
\autoref{tb:security} summarizes the results, using the same classification scheme proposed in~\cite{canella2018systematic}.
The experiment results show that \sys was able to prevent all information leaks.
%we conclude that \sys can achieve its design goals and completely defeat Spectre attacks, when incorporated with defenses against Spectre-PHT.

% \usepackage{multirow}
% \usepackage{graphicx}
\begin{table}[ht]
\centering
\caption{Empirical security evaluation of \sys.
%* means \sys can partially defeat such attacks on its own, but provide complete defenses when incorporated with Spectre-PHT defenses.
}
\label{tb:security}
\resizebox{0.49\textwidth}{!}{%
\begin{tabular}{cccc}
\multicolumn{1}{l}{}        & \multicolumn{1}{l}{}                    & in-place & out-of-place \\ \hline
% \multirow{2}{*}{Spectre-PHT} & Cross-address-space                     & \ding{51}*  & \ding{51}*              \\ \cline{2-4} 
%                             & Same-address-space                      & \ding{51}*   & \ding{51}*             \\ \hline
\multirow{2}{*}{Spectre-BTB} & Cross-address-space                     & \ding{51}   & \ding{51}            \\ \cline{2-4} 
                            & Same-address-space                      & \ding{51}   & \ding{51}             \\ \hline
\multirow{2}{*}{Spectre-RSB} & Cross-address-space                     & \ding{51}   & \ding{51}             \\ \cline{2-4} 
                            & \multicolumn{1}{l}{Same-address-space}  & \ding{51}   & \ding{51}             \\ \hline
\multirow{2}{*}{SmotherSpecter} & Cross-address-space                     & \ding{51}   & \ding{51}             \\ \cline{2-4} 
                            & \multicolumn{1}{l}{Same-address-space}  & \ding{51}   & \ding{51}             \\ \hline
\end{tabular}%
}
\end{table}

\subsubsection{Impact of CFG precision}
To study the difference between coarse-grained CFI (e.g., Inte CET~\cite{intel-cet}) and fine-grained CFI (e.g., \sys)  against BTB injection attacks, we used the SMoTherSpectre~\cite{bhattacharyya2019smotherspectre} for a demonstration. In this scenario, the attacker has to find a BTI gadget in the victim process which loads a secret in a register and terminates by an indirect branch to be able to perform BTB injection. By poisoning the BTB, the attacker transfers control to a SMoTHer Gadget to leak the secret. The SMoTHer Gadget starts with a comparison based on the target register followed by a conditional jump which enables SMoTherSpectre to leak the secret through a port contention side-channel.
\autoref{listing:cfb_attack} compares the required SMoTHer gadgets and feasibility of the attack under coarse-grained and fine-grained CFI.

% \begin{figure}[h]
% \centering
% \begin{minted}[fontsize=\footnotesize]{c}
% 1   Void bar(*char, int); /*Smother Gadget*/
% 2   Void baz(*char); /*Smother free*/
% 3    int main(int, int){
% 4        fnptr1 = baz;
% 5        fnptr2 = bar; 
% 6        fnptr1(secret, 0); /*Smother BTI*/
% 7        return 0;
% 8    }
% \end{minted}
% \captionof{listing}{SMoTherSpectre victim code}
% \label{listing:cfbending}
% \end{figure}

% \autoref{listing:cfbending} shows an example of vulnerable code to SMoTherSpectre. Assume, the code contains two functions which only the \emph{bar} function has the SMoTher Gadget. Also, assume that the code contains only one vulnerable address taken indirect branch to Smother free function \emph{baz}. 
% Without applying \sys principles, attackers can easily train the BTB and force the victim to jump to any arbitrary location including any SMoTHer Gadget.
% \autoref{listing:cfb_attack} shows the possibility of SMoTherSpectre attack in the presence of two implementation of \sys.

\begin{figure*}[h]
\begin{center}
\centering
\begin{subfigure}{0.48\textwidth}

\begin{tabular}{p{4cm}  c}
\begin{minipage}[t]{4cm}
%\begin{minted}[fontsize=\footnotesize]{gas}
\begin{lstlisting}[language=myasm]
Train_BTB:		
 0x1:mov rax, 0x20
 0x2:call *rax
foo:
  0x10: endbr64
  0x11: nop
\end{lstlisting}
\end{minipage}
&
\begin{minipage}[t]{4cm}
%\begin{minted}[fontsize=\footnotesize]{gas}
\begin{lstlisting}[language=myasm]
main: //BTI gadget
  0x0:mov rdx,[secret] 
  0x1:mov rax,0x10
  0x2:call *rax //baz()
baz: //Smother free
  0x10: endbr64
        ...
  0x14: nop
bar://Smother Gadget
  0x20:endbr64	
  0x24:cmp $0, rdx
  0x25:je <>
  
\end{lstlisting}
\end{minipage}
\\
\centering \footnotesize{\textbf{\texttt{Attacker}}} & \footnotesize{\textbf{\texttt{Victim}}}
\\
\end{tabular}
\caption{Coarse-grained enforcement of CFI (e.g. CET)}
\label{listing:cfb_attack-a}
\end{subfigure}
\begin{subfigure}{0.48\textwidth}
\begin{tabular}{p{3.2cm}  c}

\begin{minipage}[t]{4 cm}
%\begin{minted}[fontsize=\footnotesize]{gas}
\begin{lstlisting}[language=myasm]
Train_BTB:		
 0x1:mov rax, 0x20
 0x2:call *rax, L1
foo:
  0x20: cfi_lbl, L1
  0x21: nop
\end{lstlisting}
\end{minipage}
&
\begin{minipage}[t]{4.5cm}
%\begin{minted}[fontsize=\footnotesize]{gas}
\begin{lstlisting}
main: //BTI gadget
  0x0:mov rdx,[secret] 
  0x1:mov rax,0x10
  0x2:call *rax, L1//baz()
baz: //Smother free
  0x10: cfi_lbl, L1
        ...
  0x14: nop
bar://Smother Gadget
  0x20:cfi_lbl, L2	
  0x21:cmp $0, rdx
  0x22:je <>
\end{lstlisting}
\end{minipage}
\\
\centering \footnotesize{\textbf{\texttt{Attacker}}} & \footnotesize{\textbf{\texttt{Victim}}}
\\
\end{tabular}
\caption{ Fine-grained enforcement of CFI (e.g, \sys)}
\label{listing:cfb_attack-b}
\end{subfigure}

\end{center}
\caption{Speculative control-flow bending attack example.}
\label{listing:cfb_attack}
\end{figure*}

% In~\ref{listing:cfb_attack-a} attacker tries to redirect the victim's control flow to the SMoTHer gadget in branch location \texttt{0x1}. Given that the coarse-grained CFI consider the same label for all indirect forward edge branches, the attacker can successfully launch the attack.    
% Although, enforcing the Intel CET implementation of \sys principle, restricted the attacker to select her Smother gadget from a smaller set of gadgets, but as shown in above example, the attack is still possible. We study the possibility of finding SMoTHer gadgets for standard Library in existence of Intel CET in~\ref{tbl:gadgets}.

% Figure~\ref{listing:cfb_attack-b} shows the same attack when a fine-grained CFI is applied. The attacker is free to choose any arbitrary label to inject an entry to the BTB. In this example if she choose \emph{L1} he can only transfer the control to the \emph{baz} function which is safe. If she select \emph{L2}, when the victim redirected to \emph{bar} the CFI check will violated because the call's label (e.g, L1) does not match. So, in neither way she is successful to launch the attack.

\begin{table}[t]
\centering
\caption{Available SMother Gadgets in Standard Libraries}
\label{tbl:gadgets}
\resizebox{0.4\textwidth}{!}{
\begin{tabular}{ccc} 
\hline
\multirow{2}{*}{Standard Libraries} & \multicolumn{2}{c}{CFI Implementation}           \\ \cline{2-3}
                              & \textit{Coarse-grained} & \textit{Fine-grained}  \\ 
\hline
                glibc-2.29&    314   &1          \\
                libssl-1.1&    21    & 1        \\
                libcrypto-1.1 & 98    &  4      \\
                ld-2.29&     64  &    0  \\
                libstdc++&   47   &   0    \\

\hline
\end{tabular}
}
\vspace{-1.5em}
\end{table}

\autoref{tbl:gadgets} shows the number of available SMoTher Gadgets from several standard libraries. Using the constraints for the SMother Gadget identified by Bhattacharyya et al.~\cite{bhattacharyya2019smotherspectre}, we scanned for valid SMoTHer gadgets in the first 70 instructions after label instructions (\texttt{endbr64} and \texttt{cfi\_lbl}). For \sys, we used a function signature based approach for generating labels~\cite{niu14a,niu15}. As we can see, although fine-grained CFI still permits some gadgets, the number is much smaller than that available under coarse-grained CFI.

%Given that all the targets in Intel CET implementation take the same label, all the reported gadget would be available for the attacker, depends on linked standard libraries in the victim process. The number of available gadgets for fine-grained approaches depends on the accuracy of analysis for generating the labels. Using the signature based approach for generating labels in \sys, we did not find any matching label with vulnerable \emph{do\_cipher} function in OpenSSL as reported in the table~\ref{tbl:gadgets}.   

It is worth mentioning that we only use SMoTHer gadget constraints as an example of  practical gadgets. There are no clear systematic approaches to locate generic Spectre gadgets that are exploitable in practice, further analysis is required in order to find more specific constraints.  We hope to pursue this question in our future work.

\section{Related Work}
\label{sec:related}

% \usepackage{graphicx}

\begin{table*}[t]
\centering
\caption{Spectre defenses and the attacks they mitigate. Symbols show if an attack is mitigated (\CIRCLE), not mitigated (\Circle), or partially mitigated (\LEFTcircle).}
\label{tb:defense}
\resizebox{\textwidth}{!}{%
\begin{minipage}{1.25\textwidth}

\begin{tabular}{lcclccccccclc}
\multicolumn{1}{c}{\multirow{2}{*}{\textbf{Attacks}}} & \multicolumn{2}{c}{\textbf{Side-channel prevention}} &  & \multicolumn{8}{c}{\textbf{Speculation prevention}} \\ \cline{2-3} \cline{5-12} 
\multicolumn{1}{c}{} & DAWG~\cite{kiriansky18dawg} & \begin{tabular}[c]{@{}c@{}}SafeSpec~\cite{safespec}, \\ InvisiSpec~\cite{yan2018invisispec}\end{tabular} & \multicolumn{1}{c}{} & \begin{tabular}[c]{@{}c@{}}LFENCE\\ ~\cite{intelanalysis,arm_css,amd}\end{tabular} & \begin{tabular}[c]{@{}c@{}}IRBS, IBPB,\\ STIBP~\cite{intel_mitigations,amd}\end{tabular} & \begin{tabular}[c]{@{}c@{}}(SLH)~\cite{slh}, \\ (YSNB)~\cite{oleksenko2018you}\end{tabular} & Retpoline~\cite{retpoline} & RSB Stuffing~\cite{rsb_refill} & CSF~\cite{taram_csf19} & ConTExT~\cite{schwarz2019context} & \multicolumn{1}{l}{\sys} \\ \cline{2-12} 
Spectre-PHT & \LEFTcircle & \LEFTcircle &  & \CIRCLE & \Circle & \CIRCLE & \Circle & \Circle & \CIRCLE & \LEFTcircle &  \CIRCLE \footnote{Combined with any Spectre-PHT defense} \\ \hline
Spectre-BTB & \LEFTcircle & \LEFTcircle &  & \Circle & \CIRCLE & \Circle & \CIRCLE & \Circle & \LEFTcircle & \LEFTcircle & \CIRCLE  \\ \hline
Spectre-RSB & \LEFTcircle & \LEFTcircle &  & \Circle & \Circle & \Circle & \Circle & \CIRCLE & \LEFTcircle & \LEFTcircle & \CIRCLE  \\ \hline
SmotherSpectre & \Circle & \Circle &  & \Circle & \Circle & \Circle & \LEFTcircle & \LEFTcircle & \LEFTcircle & \LEFTcircle & \CIRCLE \\ \hline
\end{tabular}%
\end{minipage}

}\vspace{-1em}
\end{table*}

% \usepackage{multirow}
% \usepackage{graphicx}

\begin{table*}[t]
\centering
\caption{Spectre defenses and their overhead in terms of hardware complexity, software modification, and performance. Symbols show if overhead is high ($\uparrow$), low ($\downarrow$), or no overhead (-). The performance overhead results are based on what was reported in the studies; Please note that these values are not based on experiments on identical benchmarks and are only reported to provide a general sense of performance. } %which are based on real world usage or a specific benchmark (may not represent real world usage).}
\label{tb:overhead}
\resizebox{\textwidth}{!}{
\begin{tabular}{lcclcccccccc}
\multicolumn{1}{c}{\multirow{2}{*}{\textbf{Overhead}}} & \multicolumn{2}{c}{\textbf{Side-channel prevention}} &  & \multicolumn{8}{c}{\textbf{Speculation prevention}} \\ \cline{2-3} \cline{5-12} 
\multicolumn{1}{c}{} & DAWG~\cite{kiriansky18dawg} & \begin{tabular}[c]{@{}c@{}}SafeSpec~\cite{safespec}, \\ InvisiSpec~\cite{yan2018invisispec}\end{tabular} & \multicolumn{1}{c}{} & \begin{tabular}[c]{@{}c@{}}LFENCE\\ ~\cite{intelanalysis,arm_css,amd}\end{tabular} & \begin{tabular}[c]{@{}c@{}}IRBS, IBPB, \\ STIBP~\cite{intel_mitigations,amd}\end{tabular} & \begin{tabular}[c]{@{}c@{}}(SLH)~\cite{slh}, \\ (YSNB)~\cite{oleksenko2018you}\end{tabular} & Retpoline~\cite{retpoline} & RSB Stuffing~\cite{rsb_refill} & CSF~\cite{taram_csf19} & ConTExT~\cite{schwarz2019context} & \sys \\ \cline{2-12} 
Hardware & $\uparrow$ & $\uparrow$ &  & -- & -- & -- & -- & -- & $\downarrow$ & $\downarrow$ & $\downarrow$ \\ \hline
Software modification & -- & -- &  & $\uparrow$ & -- & $\uparrow$ & $\downarrow$ & $\downarrow$ & $\downarrow$ & $\downarrow$ & $\downarrow$ \\ \hline
Performance & 1 - 5 \% & \begin{tabular}[c]{@{}c@{}}SafeSpec: -3\%\\ InvisiSpec: 22\%\end{tabular} &  & 62 - 74.8 \% & 20 - 50 \% & \begin{tabular}[c]{@{}c@{}}SLH: 29 - 36.4 \%\\ YSNB: 60 \%\end{tabular} & 5 - 10 \% & $\downarrow$ & 2.7 - 15.2 \% & 1 - 71.14 \% & 1.9 \% \\ \hline
\end{tabular}%
}\vspace{-1em}
\end{table*}

Since the initial announcement of Spectre and Meltdown in January of 2018, several Spectre variants have appeared~\cite{spectre,kiriansky2018speculative,koruyeh2018spectre,maisuradze2018ret2spec,evtyushkin-18, ssp2018}. Spectre attacks are characterized by manipulating the prediction mechanisms to trigger speculation to an attacker chosen gadget.  They differ in what they exploit to trigger speculation: branch direction predictor (variant 1, variant 1.1)~\cite{spectre,evtyushkin-18,kiriansky2018speculative}, branch target predictor (or branch target buffer) for variant 2~\cite{spectre}, return stack buffer for Spectre-RSB (also called variant 5)~\cite{koruyeh2018spectre,maisuradze2018ret2spec}, or load-store aliasing predictor for variant 4~\cite{ssp2018}. To mitigate these attacks, several software and hardware defenses ranging from programming guidelines for cryptographic software developers~\cite{programming_guidelines} to architectural changes~\cite{safespec,yan2018invisispec} have been proposed. In this section, we will overview these defenses categorized into the Spectre attack variants that they defend against.
%
%Another way to categorize the defenses is by the speculation attack ingredient it is trying to mitigate; (1) side-channel prevention: defences developed to prevent the leakage of secret data during speculative execution. (2) speculation prevention: defences developed to abort speculation when it is dangerous.
\autoref{tb:defense} shows the Spectre attacks defenses and which attacks they mitigate and~\autoref{tb:overhead} shows the Spectre attacks defenses and their impact on hardware complexity, software modifications, and performance. \sys is the only defense that provides complete protection against all Spectre attacks with little impact on performance and implementation  overhead. 
Note that we are not considering Meltdown style attacks~\cite{meltdown, van2018foreshadow, Schwarz2019ZombieLoad, ridl, fallout, schwarz2019store} since they rely on speculation within a single instruction and therefore do not rely on manipulating the branch prediction structures.

\subsection{Spectre-PHT Defenses}
Spectre-PHT exploits the directional predictor (also called the Pattern History Table or PHT) to perform the attack. To defend against this attack, Intel, AMD, and ARM proposed to use instructions that serialize the execution (e.g. \texttt{lfence}) to stop speculation around conditional branches~\cite{intelanalysis,arm_css,amd}. Although aggressive serialization (e.g., at every branch instruction) can mitigate Spectre-PHT, it  hurts performance substantially~\cite{intelanalysis}: serializing all branch instructions will eliminate the performance benefit of the branch predictor (e.g., up to 10x slowdown~\cite{oleksenko2018you}).  Therefore, multiple proposals tried to reduce the number of serialization points introduced using static analysis to identify and serialize exploitable gadgets only~\cite{intelanalysis,spectre-msvc,Respectre,oo7}. However, these approaches miss some of the gadgets that can be exploited~\cite{Paul_spectre}. Another weakness of these defenses is that even though they stop speculative execution around exploitable gadgets, they do not stop speculative code fetches and other micro-architectural behaviors before execution (e.g., instruction cache and iTLB fills) which can also leak data~\cite{netspectre}. 

Speculative Load Hardening (SLH)~\cite{slh} and You Shall Not Bypass (YSNB)~\cite{oleksenko2018you} try to reduce the high overhead by identifying Spectre gadgets, then injecting artificial dependencies between branches and these identified gadgets.  Although this results in performance advantages over liberal fencing, they still have 36\%-60\% performance overhead~\cite{taram_csf19}. 
%SLoth~\cite{kiriansky2018speculative} is a group of micro-architectural defenses that constrain store-to-load forwarding to defend against Spectre v1.1 and v4; (1) SLoth Bear: microcode update that prevent store-to-load forwarding from either speculative stores or to speculative loads. (2) SLoth: choose candidates for forwarding based on compiler marking instructions. (3) Arctic SLoth: apply dynamic detection of load and store pairs to determine candidates for forwarding. Nevertheless, this approach does not defend against Spectre v1 and require software, compiler, and hardware changes that results in performance overhead and implementation complexity.  
Context-Sensitive Fencing (CSF)~\cite{taram_csf19} is a micro-code mitigation technique where serialization instructions are added dynamically based on run-time conditions that identify potential exploit execution.  Although CSF focuses primarily on Spectre-PHT,  the authors propose to defend against Spectre-BTB and Spectre-RSB using a special fence that would flush the BTB/RSB when transferring control to higher domains. However, flushing BTB and RSB  hurts performance since it results in mis-predictions. In addition, in a simultaneous multithreading (SMT) processor, flushing the BTB/RSB after control transfer is not enough to protect against Spectre-BTB and Spectre-RSB since structures can be polluted after a control transfer using other threads. 

\subsection{Spectre-BTB and Spectre-RSB Defenses}

Spectre-BTB exploits the BTB and Spectre-RSB exploit the RSB to perform the attack. Google proposed Return Trampoline (retpoline)~\cite{retpoline} as a software mitigation technique that defends against Spectre-BTB by replacing indirect branches with push+return instruction sequences that prevent BTB poisoning. However, this solution has high performance overhead since it stops speculation (similar to serialization). In addition, it can be bypassed using \emph{ret} instructions since they cause mis-speculation through the BTB in some processors (e.g., Intel's core i7 processors starting from Skylake).  In particular, those processors predict the address of a \emph{ret} instruction from the BTB when the RSB is empty (which can be forced by executing unmatched returns). To solve this exploit, RSB stuffing~\cite{rsb_refill} was proposed to intentionally fill the RSB with the address of a benign delay gadget to avoid misspeculation on context switches. Although this technique can partially mitigate Spectre-BTB (when using \emph{ret} to trigger speculation through BTB), it can also defend against SpectreRSB cross-domains attack. However, since we are filling the RSB on context switch, stored entries for the currently running process will be lost when execution is switched back to the current process (i.e. performance loss due to losing speculation information). In contrast, \sys saves committed RSB entries per process in case of a context switch out of the process and restores them when execution returns to the process, which results in improving the prediction performance of \emph{ret} instructions.

Intel and AMD added new instructions to their instruction set architecture (ISA) that can control indirect branches to defend against Spectre-BTB~\cite{intel_mitigations,amd}. The addition consists of three controls:

\begin{itemize}
    \item Indirect Branch Restricted Speculation (IBRS): allows processors to enter IBRS mode (privileged mode) and execute indirect branches that are not influenced by less privileged mode.  
    \item Single Thread Indirect Branch Prediction (STIBP): will not allow a hyperthread running on a core to use branch predictor entries inserted by the other thread running on the same core.  
    \item The Indirect Branch Predictor Barrier (IBPB): allows processors to flush BTB and clear their state. This way, the code executed before the barrier cannot impact branch prediction of the code executed after this instruction. 
\end{itemize}

These new ISA instructions defend only against Spectre-BTB. In addition, they have a high performance overhead; up to 24\% on Skylake and up to 53\% on Haswell~\cite{intel_microcode_overhead}.

\subsection{Spectre All Variants Defenses}

Several mitigations were proposed to defend against all variants of Spectre. Dynamically Allocated Way Guard (DAWG)~\cite{kiriansky18dawg,nomo} was proposed to provide isolation between protection domains by partitioning the cache at the cache way granularity. Although this method can prevent leakage of the data through a cache side-channel, it requires domains enforcement management in the software, defending cache as leakage source only, and it can not protect against attacks that are performed within the same address space or isolation domain. In addition, since it is a cache specific defense, other micro-architectural structures can be used for communication (e.g. branch predictor).

SafeSpec~\cite{safespec} and InvisiSpec~\cite{yan2018invisispec} are hardware mitigation techniques that are similar to DAWG in the way that they are both trying to prevent side-channel communication from speculative instructions. Therefore, they propose to mitigate the side-effect of speculative execution on the micro-architectural state; shadow micro-architectural structures for caches and Translation Lookaside Buffers (TLBs) were added to store transient effect of speculative instructions. These effects will be committed to caches and TLBs only when speculation is deemed correct and flush the changes from the shadow structures otherwise. Although these solutions outperform software solutions, they require making disruptive changes to the processor/memory architecture and consistency models. 

ConTExT~\cite{schwarz2019context} introduced protection for secret data from speculative execution. Specifically, the technique introduces a new memory mapping (called non-transient mapping) which tracks data that must not be accessed by speculative instructions. Nevertheless, this solution requires changes to the architecture and the operating system, and developer involvement to annotate the secret data.  It also incurs high performance overhead for security-critical applications.

\section{Concluding Remarks}\label{sec:conclude}

In this paper, we presented a new defense that protects speculative processors against misspeculation targeting the branch target buffer (BTB) and the return stack buffer (RSB).  These attacks are arguably the most dangerous speculation attacks because they can bypass compiler inserted fences.   Prior defenses either excluded these attacks from their threat model, or implemented aggressive limits to speculation that dramatically degraded performance.  In contrast, \sys provides complete protection against these dangerous attacks, with little impact on performance, and with minimal hardware complexity.

\sys introduces the idea of using CFI, explored previously as a protection against control-flow hijacking attacks for committed instructions (i.e., even on non-speculative processors), as a defense against speculation attacks.  In particular, \sys verifies the forward-edge of CFI on the instructions in the speculative path and only allows speculation if  CFI labels match protecting against Spectre-BTB attacks.  It also verifies the backward-edge using a unified shadow call stack, protecting against Spectre-RSB attacks.  Essentially, \sys moves the CFI check to the decode stage of the pipeline, preventing speculative execution of instructions unless they conform to the CFI annotations.  For normal programs, this results in negligible performance degradation since it only prevents speculation with mismatching CFI labels, which will most likely result in misspeculation.  By stopping misspeculation, we benefit from avoiding cache pollution and other resource waste during misspeculation.

Combined with recent proposals to mitigate Spectre-PHT, we believe \sys mitigates the threat from known speculation attacks.  Moreover, it does so without sacrificing performance due to speculative execution and with minimal modifications to the processor pipeline.

\section*{Acknowledgements}

This paper was partially supported by NPRP grant 8-1474-2-626 from the Qatar National Research Fund (a member of Qatar Foundation). The statements in the paper are solely the responsibility of the authors.

%\newpage

%\section*{Acknowledgment}
\bibliographystyle{plain}
\bibliography{ref}

\vspace{12pt}
\color{red}
%IEEE conference templates contain guidance text for composing and formatting conference papers. Please ensure that all template text is removed from your conference paper prior to submission to the conference. Failure to remove the template text from your paper may result in your paper not being published.

\end{document}